\documentclass[11pt,a4paper]{article}
\usepackage{jheppub}
\usepackage{slashed}
\usepackage{arydshln}
\usepackage{mathrsfs}
\usepackage{dsfont}
\usepackage{braket}
\usepackage{bbm}
\usepackage{slashed}
\usepackage{bm}

\pdfoutput=1

\newcommand\beal{\begin{align}}

\newcommand\nn{\nonumber}

\newcommand{\eq}[1]{\begin{equation}#1\end{equation}}
\newcommand{\al}[1]{\begin{align}#1\end{align}}

\newcommand{\spl}[1]{\begin{split}#1\end{split}}
\newcommand{\be}{\begin{equation}}
\newcommand{\ee}{\end{equation}}
\newcommand{\bea}{\setlength\arraycolsep{2pt} \begin{eqnarray}}
\newcommand{\eea}{\end{eqnarray}}
\newcommand{\w}[1]{\\[0.#1cm]}
\def\cD{{\cal D}}

\def\be{\begin{eqnarray}}
\def\ee{\end{eqnarray}}
\def\ba{\begin{array}}
\def\ea{\end{array}}
\def\bec{\begin{center}}
\def\ec{\end{center}}

\def\cN{{\cal N}}
\def\ft#1#2{{\textstyle{{\scriptstyle #1} \over {\scriptstyle #2}}}}

\title{One Loop Tests of Supersymmetric Higher Spin $\bm{ {\rm AdS}_4/{\rm CFT}_3} $}
\author[1]{Yi Pang,}
\author[2]{Ergin Sezgin}
\author[2]{and Yaodong Zhu}
\affiliation[a]{Max-Planck-Insitut f\"{u}r Gravitationsphysik (Albert-Einstein-Institut)
Am M\"{u}hlenberg 1, DE-14476 Potsdam, Germany}
\affiliation[b]{George and Cynthia Woods Mitchell Institute for Fundamental Physics and Astronomy,\\
Texas A\&M University, College Station, TX 77843, USA}
\emailAdd{yi.pang@aei.mpg.de}
\emailAdd{sezgin@physics.tamu.edu}
\emailAdd{yaodongmatt@physics.tamu.edu}
\abstract{We compute one loop free energy for $D=4$ Vasiliev higher spin gravities based on Konstein-Vasiliev algebras $hu(m;n|4)$, $ho(m;n|4)$ or $husp(m;n|4)$ and subject to higher spin preserving boundary conditions, which are conjectured to be dual to  the $U(N)$, $O(N)$ or $USp(N)$ singlet sectors, respectively, of free CFTs on the boundary of $AdS_4$.
Ordinary supersymmetric higher spin theories appear as special cases of Konstein-Vasiliev theories, 
when the corresponding higher spin algebra contains $OSp({\cal N}|4)$ as subalgebra. In $AdS_4$ with $S^3$ boundary, we use a modified spectral zeta function 
method, which avoids the ambiguity arising from summing over infinite number of spins. We find that the 
contribution of the infinite tower of bulk fermions vanishes. As a result, the free energy is the sum of those which arise in type A and type B models with internal symmetries, the known mismatch between the bulk and boundary free energies for type B model persists, and ordinary supersymmetric higher spin theories exhibit the mismatch as well. The only models that have a match are type A models with internal symmetries, corresponding to $n=0$.
The matching requires  identification of the inverse Newton's constant $G_N^{-1}$  with $N$ plus a proper integer as was found previously for special cases. In $AdS_4$ with $S^1\times S^2$ boundary,  the bulk one loop free energies match  those of the dual free CFTs  for arbitrary $m$ and $n$. 
We also show that a supersymmetric double-trace deformation of free CFT based on $OSp(1|4)$ 
does not contribute to the ${\cal O}(N^0)$ free energy, as expected from the bulk.
}

\begin{document}

\maketitle

\flushbottom

\section{Introduction}

It has been known for sometime that the conjectured holographic duals of higher spin (HS) gravities \cite{Vasiliev:1992av} can  be as simple as free CFTs  living on the boundary of anti-de Sitter spacetime. Moreover, it has also  been  noted that the duality is expected to arise in weakly coupled  regimes of both bulk and boundary field theories. Therefore, one expects that higher spin AdS/CFT correspondence should be amenable to test order by order in perturbation theory.  

Free CFTs arise in conjectured dualities in the context of parity invariant 
HS gravities in 4D subject to HS symmetry preserving boundary conditions. There are
two types of parity invariant Vasiliev HS gravities, known as type A and B \cite{Sezgin:2003pt}. In their simplest forms, they both contain an infinite tower of massless even spin fields, each occurring once. They differ  from each other in the parity of the spin-0 field, which is parity even (odd) in type A (B) theory. It has been  conjectured that type A theory with  $\Delta=1$ boundary condition imposed on  the  scalar is dual to the $O(N)$  singlet sector  of $N$ free real scalars \cite{Klebanov:2002ja}, while type B  theory with $\Delta=2$ boundary condition  imposed on the pseudoscalar is dual to the $O(N)$ singlet sector 
of $N$ free  Majorana fermions \cite{Sezgin:2003pt} (for earlier work in which HS holography involving CFTs with matrix valued free fields, see  \cite{Sezgin:2002rt}). These are HS symmetry preserving boundary conditions,  with standard boundary conditions imposed on all other fields understood. 
The dual CFT can be altered by changing the boundary conditions imposed on the spin-0 field in such a way  that they break HS symmetry.  For instance, type A model with $\Delta=2$ boundary condition on the scalar  is conjectured to be dual to the critical $O(N)$  vector model \cite{Klebanov:2002ja}, while type B model with $\Delta=1$ boundary condition  imposed  on the pseudoscalar is conjectured to be dual to $O(N)$  Gross-Neveu model \cite{Sezgin:2003pt}. 

An important test of the holography is to match the free energy of the bulk theory with 
that of the CFT defined on the conformal boundary of the bulk geometry. Assuming the 
bulk HS theory possesses an action formulation, the partition function evaluated on 
Euclidean $AdS_4$ can be expanded in terms of $G_N$ as
\be
 F_{\rm bulk}=\frac1{G_N}F^{(0)}_{\rm bulk}+F^{(1)}_{\rm bulk}+G_NF^{(2)}_{\rm bulk}+\cdots
 \label{Fbulk}\ .
 \ee
When the bulk Euclidean $AdS_4$ is the hyperbolic space $H_{4}$ whose conformal 
boundary is a round $S^3$, the free energy of  the bulk HS theory should 
match with that of a free CFT on a round $S^3$. The free energy of a free CFT on $S^3$ 
takes the simple form \cite{Klebanov:2011gs}
\be
F_{\rm CFT}=NF^{(0)}_{\rm CFT}\ ,
\ee
where $F^{(0)}_{CFT}$ is the free energy of a single component in $U(N)$ or $O(N)$ vector model. 
The zeroth-order contribution $F^{(0)}_{\rm bulk}$ has not been computed so far due to the lack of an action for  
Vasiliev theory with all the required properties. We will return to this point in the conclusions.
Matching $F_{\rm bulk}$ with $F_{\rm CFT}$ necessarily requires that $F_{\rm bulk}$ is proportional to $F^{(0)}_{\rm CFT}$ at each order in the small  $G_N$ expansion and that $G_N$ is identified in terms of $N$ as
\be
G^{-1}_{N}\rightarrow\gamma(N+\Delta N)\ , 
\label{renorm}
\ee
with $\gamma$ and $\Delta N$ being constants, and $\Delta N$ should be a fixed integer for a given bulk/boundary dual pair. Therefore, the higher order quantum affects simply the relation  between $G_N$ and $N$. Assuming Fronsdal type quadratic action for the massless HS fields, one loop computations have shown that 
these requirements  are fulfilled in the conjectured  duality between type A theory and the  bosonic 
$O(N)$ vector model \cite{Giombi:2013fka}. However, for the conjectured 
duality between type B theory  and the fermionic $O(N)$ vector model \cite{Sezgin:2003pt}, 
these requirements are not satisfied since  $F^{(1)}_{\rm bulk}$ and 
$F^{(0)}_{\rm CFT}$ are not proportional to each other. 
Matching of free energy was also found in the  type A/critical $O(N)$ 
vector duality, but not in the type B/$O(N)$ Gross-Neveu duality. In critical $O(N)$ vector model, the conformal dimensions of HS currents receive quantum corrections. The leading $1/N$ corrections are summarized in \cite{Skvortsov:2015pea}. These anomalous dimensions of HS currents at ${\cal O}(1/N)$ should be compared with the one loop corrections to the $AdS$ energies of HS fields computed directly from the bulk HS theory. It would be interesting to check whether they match precisely.

The principal aim  of this paper is to extend the one loop tests by computing the free energies in a wider class of HS theories in 4D that are expected to be dual to free CFTs on the boundary of $AdS_4$. In particular, we wish to study the consequences of supersymmetry which combine type A and type B spectra of fields with an infinite tower of massless fermions. The underlying HS algebras, denoted by $hu(m;n|4)$, $ho(m;n|4)$ and $husp(m;n|4)$, and their representations were determined sometime ago by Konstein and Vasiliev \cite{Konstein:1989ij}. These representations are obtained from two-fold tensor products of bosonic and fermionic singleton representations of $SO(3,2)$ which also carry fundamental representations of classical Lie groups. Vasiliev equations for these theories are described in detail in \cite{Sezgin:2012ag}. Their spectral properties will be summarized in the next section. Suffices to mention here that generically their underlying HS algebras serve as infinite dimensional supersymmetry algebras, and only in special cases, namely when $m=n=2^k$ for some $k$ corresponds to the fundamental spinor representation of $O({\cal N})$, they contain the $AdS_4$ superalgebra $OSp({\cal N}|4)$, in which case the singletons in the boundary CFT are in the spinor representations of the $R$-symmetry group $SO({\cal N})$
\footnote{In order to distinguish the notion of supersymmetry in generic Konstein-Vasiliev models versus the special cases where $OSp({\cal N}|4)$ arises as a subalgebra, we shall sometimes refer to the latter ones as ``ordinary supersymmetric HS theories".}.
We shall also consider extension of these models by introduction of internal symmetry \cite{Sezgin:2012ag}.

When the boundary of $AdS_4$ is $S^3$, we compute the one loop free energy by using the modified spectral zeta function method, which avoids the ambiguity arising from summing over infinite number of spins.  As a side result, we obtain the contributions of the even and odd spin towers of HS fields separately. Furthermore we find that the contribution of the infinite tower of fermionic fields to the free energy vanishes. Putting all results together, we find that the bulk free energy may match that of the dual free CFT only for type A models. Their spectrum consist of bosonic fields arising from the tensor product of two bosonic singletons in fundamental representation of classical Lie groups. The matching requires  identification of the inverse Newton's constant $G_N^{-1}$  with $N$ plus a proper
integer as was found previously for special cases. Note that mismatch in the free energy at one loop occurs in particular for type B models whose spectrum consists of bosonic fields arising from the tensor product of two spinor singletons in fundamental representation of classical Lie groups.

When $AdS_4$ is written in the  thermal $AdS$ coordinates, with the boundary being  $S^1\times S^2$, we find that the bulk one loop free energies match those of the dual free CFTs for generic Konstein-Vasiliev models.

The ${\cal N}=1$ higher spin theory admits ${\cal N}=1$ mixed boundary condition which corresponds to adding a supersymmetric double-trace deformation in the free CFT. We show that such a double-trace  deformation does not contribute to the ${\cal O}(N^0)$ free energy, compatible with the fact that imposing  mixed boundary condition does not change the bulk spectrum and therefore the bulk one loop free energy  remains the same.

The rest of the paper is organized as follows. In Section 2 we review the spectra of HS gravities based on HS algebras $hu(m;n|4)$, $ho(m;n|4)$ and $husp(m;n|4)$. In Section 3, we compute the one loop free energies of these theories in $AdS_4$ with $S^3$ boundary, where we also consider the ordinary supersymmetric HS theories with internal symmetry. We adopt an alternate regularization scheme introduced in \cite{Bae:2016rgm} in the bosonic sector, then generalize the method also to the fermionic sector. In Section 4, we compare the results obtained in the bulk with the corresponding ones in the boundary CFTs. In Section 5, we implement the one loop test to HS theories in thermal AdS with the dual CFTs on boundary $S^1\times S^2$. In Section 6 we study a possible mixed boundary condition for $\mathcal{N}=1$ higher spin theory and the effect on the free energy on the CFT side where a supersymmetric double-trace deformation is turned on. We summarize and comment on our results in Section 7, and comment on possible ways to approach the problem of mismatch of free energies in type B and ordinary supersymmetric HS theories and their conjectured duals. We also comment on the action formulation proposed in \cite{Boulanger:2015kfa} in the context of classical free energy in the bulk. The validity and detailed calculation of the alternate regularization method adopted in this paper are shown in Appendix A.


\section{Konstein-Vasiliev and supersymmetric higher spin theories}


The group theoretical building blocks for the construction of the physical spectra of HS theories 
in $AdS_4$ are the singleton representations of $SO(3,2)$. There are two of them referred to as Di and Rac. 
Using the standard notation $D(E_0,s)$ for the discrete unitary representations of
$sp(4;\mathbb{R}) \sim SO(3,2)$, where $E_0$ is the lowest energy and $s$ is the spin of the
lowest weight state, Di refers to the $D(1,1/2)$ and Rac refers to the $D(1/2,0)$ representations.
An important property these representations have is given by Flato-Fronsdal theorem which states that
\bea
 {\rm Rac} \otimes {\rm Rac} &=& \sum_{s=0}^\infty D(1+s,s)\ ,
\qquad
 {\rm Di} \otimes {\rm Di} = D(2,0) + \sum_{s=1}^\infty D(1+s,s)\ ,
\nn\w2
{\rm Di} \otimes {\rm Rac} &=& \sum_{s=0}^\infty D(3/2+s,1/2+ s)\ ,
\label{ff}
\eea
where $s=0,1,2,...$. The representations $D(1+s,s)$ are massless spin $s$ fields, and $D(2,0)$ is a 
massless pseudoscalar field. To introduce internal symmetry, consider the singleton
representations
\be
S_+:=({\rm Rac},m)\oplus ({\rm Di},n)\ , \qquad S_-:=({\rm Di},m)\oplus ({\rm Rac},n)\ .
\ee
where $m$ labels the fundamental representations of $u(m)$ or $usp(m)$ or a vector
representation of $so(m)$. It has been shown that the physical spectra of three
types of HS theories, based on HS algebras denoted by
$hu(m;n|4), ho(m;n|4), husp(m;n|4)$, are obtained from the following tensor
products of the singletons
\bea
hu(m;n|4)&:& \ \ S_+\otimes \bar S_+\ ,\qquad\qquad\ \ \ \ \ \
hu(n;m|4):\qquad \ \   S_-\otimes \bar S_-\ ,
\label{singhu}\\[5pt]
ho(m;n|4)&:& \left(S_+\otimes S_+\right)_S\ ,\qquad\qquad\quad
ho(n;m|4):\quad\ \ \  \left(S_-\otimes S_-\right)_S\ ,
\label{singho}\\[5pt]
husp(m;n|4)&:& \left(S_+\otimes S_+\right)_A\ ,\qquad\quad\ \
husp(m;n|4): \qquad\left(S_-\otimes S_-\right)_A\ ,
\label{singhusp}
\eea
where $\left(\cdot\right)_S$ and $\left(\cdot\right)_A$ stand for symmetric and antisymmetric tensor products, respectively. These algebras contain $u(m)\otimes u(n)$, $o(m)\otimes o(n)$ and $usp(m) \otimes usp(n)$ as maximal bosonic subalgebras. The resulting spectra are as follows \cite{Konstein:1989ij}
\be
\begin{array}{lcll}
hu(m;n|4) &:& (m^2-1,1)\oplus (1,n^2-1)\oplus (1,1)\oplus (1,1)&s=0,1,2,3,\dots\\[5pt]
&& (m,\bar n)\oplus (\bar m,n)& s=\ft12,\ft32,\ft52,\dots
\\[10pt]
ho(m;n|4) &:& (\ft12 m(m-1),1)\oplus (1,\ft12 n(n-1))&s=1,3,\dots
\\[5pt]
&& (\ft12 m(m+1)-1,1)\oplus (1,\ft12 n(n+1)-1)\oplus (1,1)\oplus (1,1)&s=0,2,4,\dots
\\[5pt]
&& (m, n) & s=\ft12,\ft32,\ft52,\dots
\\[10pt]
husp(m;n|4) &:& (\ft12 m(m+1),1)\oplus (1,\ft12 n(n+1))&s=1,3,\dots
\\[5pt]
&& (\ft12 m(m-1)-1,1)\oplus (1,\ft12 n(n-1)-1)\oplus (1,1)\oplus (1,1)&s=0,2,4,\dots
\\[5pt]
&& (m, n) & s=\ft12,\ft32,\ft52,\dots\ ,
\end{array}
\label{KS}
\ee
where the dimensions of the representations are shown. While there are the isomorphisms $hu(m;n|4) \sim hu(n;m|4)$, $ho(m;n|4) \sim ho(n;m|4)$ and $husp(m;n|4) \sim husp(n;m|4)$, the corresponding spectra listed above form inequivalent representations since there are $\{m^2, m(m+1)/2,
m(m-1)/2\}$ scalars in $D(1,0)$ representations, and  $\{n^2, n(n+1)/2,
n(n-1)/2\}$ scalars in $D(2,0)$ representations of $SO(3,2)$, in the cases of $hu(m;n|4)$, $ho(m;n|4)$,
$husp(m;n|4)$ respectively.
The models with $mn>0$ contain fermions and are based on HS algebras that are superalgebras in the sense that they involve bosonic and fermionic generators and graded commutators. However, unless $m=n=2^{N/2-1}$ or $m=n=2^{(N-1)/2}$, these algebras do not contain a finite dimensional superalgebra and as such they are infinite dimensional algebras. In the case of $m=n=2^{N/2-1}$, the $\text{Rac}$ and $\text{Di}$ belong to left and right handed fundamental spinor representations of $SO({\cal N})$ and we have the isomorphisms
\be
\label{shs1}
shs^E({\cal N}|4)~{\cong}~\left\{\ba{ll} hu\left(2^{\frac{\cal N}{2}-1};2^{\frac{\cal N}{2}-1}\Big\vert 4\right)\  & \qquad{\cal N} = 2\ {\rm mod}\ 4 \ ,
\\[5pt]
husp\left(2^{\frac{\cal N}{2}-1};2^{\frac{\cal N}{2}-1}\Big\vert 4\right) &\qquad {\cal N} = 4\ {\rm mod}\ 8\ ,
\\[5pt]
ho\left(2^{\frac{\cal N}{2}-1};2^{\frac{\cal N}{2}-1}\Big\vert 4\right)&\qquad {\cal N} = 8\ {\rm mod}\ 8\ .
\ea
\right.
\ee
The HS superalgebra $shs^E({\cal N}|4)$ contains the ${\cal N}$ extended  $AdS_4$ superalgebra  $OSp({\cal N}|4)$ as a subalgebra. In the case of $m=n=2^{({\cal N}-1)/2}$, the $\text{Di}$ and $\text{Rac}$ belong to the $2^{({\cal N}-1)/2}$ dimensional fundamental spinor representations of $SO({\cal N})$ and we have the isomorphisms
\be
\label{shs2}
shs^E({\cal N}|4)~{\cong}~\left\{\ba{ll} ho\left(2^{({\cal N}-1)/2};2^{({\cal N}-1)/2}
\Big\vert 4\right)\  & \qquad{\cal N} = 1\ {\rm mod}\ 8 \ ,
\\[5pt]
husp\left(2^{({\cal N}-1)/2};2^{({\cal N}-1)/2}
\Big\vert 4\right) &\qquad {\cal N} = 5\ {\rm mod}\ 8\ .
\ea
\right.
\ee
As for the case of  ${\cal N}$=3 mod 4, it has been shown in \cite{Sezgin:2012ag} that it is equivalent to the case of ${\cal N}$=4 mod 4.
The $OSp({\cal N}|4)$ supermultiplet content of the spectra described above can be determined in a straightforward way but  this information is not needed for the purposes of this paper.

The supersymmetric HS models  described above can be extended by introduction of internal symmetry. In this case, the Di and Rac representations not only carry the spinor representation of $SO({\cal N})$ but also a fundamental representation of  a classical Lie algebra. Working out their tensor products yields the spectrum of the expected dual HS theory, which can be found in Table 5 of \cite{Sezgin:2012ag}.


\section{Free energies of Konstein-Vasiliev higher spin
theories in $AdS_4$ with $S^3$ boundary }


In this section we shall compute the free energy of Konstein-Vasiliev HS theories in $AdS_4$ with $S^3$ boundary, imposing the HS symmetry preserving boundary conditions.
Free energy of bosonic HS fields in $AdS_4$ has been studied in \cite{Giombi:2013fka,Giombi:2014iua,Giombi:2014yra,Giombi:2016ejx}. The regularization scheme that has been used in summing over infinite tower of HS fields, however, is very complicated. Here, we employ an alternate method which is much simpler, utilizing the character of irreducible representation of $SO(2,3)$. This method was introduced in \cite{Bae:2016rgm} to compute the one loop free energy of massive HS fields, but was not applied to the computation of the above free energies to exhibit the contributions of the infinite tower of odd and even spins separately. In what follows we shall use the alternate method to compute these contributions separately. We then generalize the method and apply it to the computation in bulk fermion sector in the subsequent subsection.

The one loop correction to the free energy is defined as $F^{(1)}=-\log Z^{(1)} $ where $Z^{(1)}$ is the one loop partition function. For HS theory with $n_S$ real scalars, $n_P$ pseudoscalars, $n_1$ copies of fields with
$s=1,3,...,\infty$, $n_2$ copies of fields with $s=2,4,...,\infty$ fields and $n_F$ copies of spin $1/2,3/2,...,\infty$ fields, we have
\bea
&& F^{(1)} (n_S,n_P,n_1,n_2,n_F)  =
 \ft12 n_S \log {\rm det}_1\, \cD_B(1,0) + \ft12 n_P \log {\rm det}_2\, \cD_B(2,0)
\nn\w2
&& +\ft12 n_1 \sum_{k=0}^\infty \Big[ \log\det\,\cD_B(2k+2,2k+1)-  \log\det\,\cD_B(2k+3,2k)\Big]
\label{master}\w2
&& +\ft12 n_2 \sum_{k=1}^\infty \Big[ \log\det\,\cD_B(2k+1,2k)-  \log\det\,\cD_B(2k+2,2k-1)\Big]
\nn\w2
&& - \ft12 n_F \log\det\, \cD_F(\ft32,\ft12)
-\ft12 n_F \sum_{k=1}^\infty \Big[ \log\det\,\cD_F (k+\ft32,k+\ft12) -  \log\det\,\cD_F(k+\ft52,k-\ft12)\Big]\ ,
\nn\label{F1}
\eea
where we have defined
\bea
{\cal D}_B(\Delta,s) &=& \left[-\nabla^2 +\Delta(\Delta-3) -s\right]\ ,
\nn\w4
{\cal D}_F(\Delta,s) &=&  \left[ -\slashed{\nabla}^2 +\Delta(\Delta-3) +\ft94 \right]\ .
\eea
The negative contributions in the bosonic sector and the positive contributions in the fermionic sector are due to ghosts. In computing ${\rm det}_1$ and ${\rm det}_2$,  the irregular ($\Delta_-=1$) and regular ($\Delta_+=2$) boundary conditions are to be used.

For a differential operator of the form ${\cal D} = -\nabla^2 + X$, or ${\cal D} = -\slashed{\nabla}^2 + Y$, writing
\be
-\log\det  {\cal D} =\int_{0}^{\infty} \frac{dt}{t} K_{\cal D}(t) \ ,\qquad K_{\cal D}(t) := {\rm Tr}\left[  e^{-t{\cal D}}\right]\ ,
\ee
and defining the spectral zeta function
\be
\zeta_{\cal D}(z) := \frac{1}{\Gamma(z)} \int_{0}^\infty dt \,t^{z-1} K_{\cal D}(t)\ ,
\ee
one finds the standard result \cite{Hawking:1976ja}
\be
-\log \det {\cal D} = \zeta_{\cal D}(0) \log (\ell^2\Lambda^2) + \zeta^\prime_{\cal D}(0)\ ,
\ee
where $\ell$ is the $AdS$ radius and $\Lambda$ is the renormalization scale. For fields of aribrary spins in hyperbolic space $H_4$, the spectral zeta function technique has been developed in \cite{Camporesi:1991nw,Camporesi:1993mz} to compute their one loop effective potentials.

\subsection{Bosons}

Upon Euclideanization of $AdS_4$ to $H_4$, the boundary is $S^3$ and in this setting
various free energies of the bosonic HS theory are given by
\bea
F^{(1)}_{{\rm even}\,1}&=&-\frac12\Big[\zeta^B_{(1,0)} (0)+\sum^{\infty}_{s=2,4,\cdots}\Big(\zeta^B_{(s + 1, s)}(0)-\zeta^B_{(s+2, s-1)}(0)\Big)\Big]\log (\ell^2\Lambda^2)
\nn\\
&&-\frac12\Big[\zeta^{B\prime}_{(1,0)} (0)+\sum^{\infty}_{s=2,4,\cdots}\Big(\zeta^{B\prime}_{(s + 1, s)}(0)-\zeta^{B\prime}_{(s+2, s-1)}(0)\Big)\Big]\ ,
\nn\\
F^{(1)}_{{\rm even}\,2}&=&-\frac12\Big[\zeta^B_{(2,0)} (0)+\sum^{\infty}_{s=2,4,\cdots}\Big(\zeta^B_{(s + 1, s)}(0)-\zeta^B_{(s+2, s-1)}(0)\Big)\Big]\log (\ell^2\Lambda^2)
\nn\\
&&-\frac12\Big[\zeta^{B\prime}_{(2,0)} (0)+\sum^{\infty}_{s=2,4,\cdots}\Big(\zeta^{B\prime}_{(s + 1, s)}(0)-\zeta^{B\prime}_{(s+2, s-1)}(0)\Big)\Big]\ ,
\nn\\
F^{(1)}_{\rm odd}&=&-\frac12\sum^{\infty}_{s=1,3,\cdots}\Big(\zeta^B_{(s + 1, s)}(0)
-\zeta^B_{(s+2, s-1)}(0)\Big)\log (\ell^2\Lambda^2)
\nn\\
&&-\frac12\sum^{\infty}_{s=1,3,\cdots}\Big(\zeta^{B\prime}_{(s + 1, s)}(0)
-\zeta^{B\prime}_{(s+2, s-1)}(0)\Big)\ ,
\eea
where $F^{(1)}_{{\rm even}\,1}$ and $F^{(1)}_{{\rm even}\, 2}$  denote the total free energy of all even spin fields $s=0,2,4\cdots$, in which the scalar satisfies $\Delta=1$ and $\Delta=2$  boundary conditions, respectively, and  $F^{(1)}_{\rm odd}$ denotes the total free energy of all odd spin fields $s=1,3,5\cdots$.

As stated earlier, we now employ a simpler method than those used previously, utilizing the character of irreducible representation of $SO(2,3)$. The method is based on the observation that the spectral zeta function of a bosonic spin-$s$ field can be recast in the form
\be
\zeta^B_{(\Delta, s)}(z)=\frac{1}{\Gamma(z)}\int^\infty_0 d\beta
\Big[ \mu(z,\beta)+\nu(z,\beta)\frac{\partial^2}{\partial\alpha^2}\Big]
\chi_{\Delta,s}(\beta,\alpha)~\Big\vert_{\alpha=0}\ ,
\label{cb2}
\ee
in which
\bea
&&\chi_{\Delta,s}(\beta,\alpha)=\frac{e^{-\beta(\Delta-\ft32)}\sin[(s+\ft12)\alpha]}{4\sinh\frac{\beta}{2}\sin\frac{\alpha}{2}(\cosh\beta-\cos\alpha)}\ ,
\nn\\
&&\mu(z,\beta)=\ft{1}{3}\sinh\ft{\beta}{2}\Big[ f_1(z,\beta) \Big(-6+ \sinh^2\ft{\beta}{2}\Big) +4 f_3(z,\beta) \sinh^2\ft{\beta}{2} \Big]\ ,
\nn\\
&&\nu(z,\beta)=-4f_1(z,\beta)\sinh^3\ft{\beta}{2}\ ,
\nn\\
&&f_n(z,\beta)=
    \sqrt{\pi}\int^\infty_0duu^n\tanh(\pi u)(\ft{\beta}{2u})^{z-\ft{1}{2}}J_{z-1/2}(u\beta)\ ,
\label{bb}
\eea
where $\chi_{\Delta,s}(\beta,\alpha)$ is the character of a representation of $SO(3,2)$ labeled by $D(\Delta,s)$. Owing to the $e^{-\beta(\Delta-\ft32)}$ factor in the character, $\sum_s \zeta_{(\Delta, s)}(z)$ is convergent. Therefore, no regularization is needed in performing the sum over infinitely many spins. This is the desired feature for computing the one loop free energy of HS theory where the summation over infinitely many spins is encountered.
It was also noticed by \cite{Bae:2016rgm} that since the one loop free energy depends only on $\zeta(0)$ and $\zeta'(0)$, an alternate zeta function $\widetilde\zeta(z)$ is physically equivalent to the original $\zeta(z)$, provided that $\widetilde\zeta(0)=\zeta(0)$, and $\widetilde\zeta'(0)=\zeta'(0)$. Thus, for the convenience of calculation, one can in fact
utilize an alternate zeta function which is physically equivalent to the original zeta function. For bosonic HS fields, one choice of the alternate zeta function takes the form \cite{Bae:2016rgm}
\be
\widetilde{\zeta}^B_{(\Delta, s)}(z)= \frac{1}{\Gamma(2z)} \int^\infty_0 d\beta\, \beta^{2z-1} \coth \ft{\beta}{2}
\Big[1+ \left(\sinh^2\ft{\beta}{2}\right)\, \partial^2_\alpha \Big]\chi_{\Delta,s}(\beta,\alpha)~\Big\vert_{\alpha=0}\ .
\label{azetab}
\ee
The physical equivalence between the alternate spectral zeta function and the original one \eqref{cb2} is shown in the appendix. The total character of all even spin fields and that of all odd spin fields are computed as
\bea
\chi_{{\rm even}\,1}(\beta,\alpha)&=&\chi_{1,0}(\beta,\alpha)
+\sum_{s=2,4,\cdots}(\chi_{s+1,s}(\beta,\alpha)-\chi_{s+2,s-1}(\beta,\alpha))
\nn\\
&=&\frac{1+\cos\alpha+\cosh\beta+\cosh2\beta}{4(\cos\alpha
-\cosh\beta)^2(\cos\alpha+\cosh\beta)}\ ,
\label{CE1}\\
\chi_{{\rm even}\,2}(\beta,\alpha)&=&\chi_{2,0}(\beta,\alpha)
+\sum_{s=2,4,\cdots}(\chi_{s+1,s}(\beta,\alpha)-\chi_{s+2,s-1}(\beta,\alpha))
\nn\\
&=&\frac{1+\cos\alpha+\cos2\alpha+\cosh\beta}{4(\cos\alpha
-\cosh\beta)^2(\cos\alpha+\cosh\beta)}\ ,
\label{CE2}\\
\chi_{\text{odd}}(\beta,\alpha)&=&\sum_{s=1,3,\cdots}(\chi_{s+1,s}(\beta,\alpha)
-\chi_{s+2,s-1}(\beta,\alpha))
\nn\\
&=&\frac{\cos\alpha+\cosh\beta+2\cos\alpha\cosh\beta}{4(\cos\alpha
-\cosh\beta)^2(\cos\alpha+\cosh\beta)}\ .
\label{CO}
\eea
Substituting the results above into \eqref{azetab}, we find
\bea
\widetilde{\zeta}^B_{\text{even,1}}(z)&=&\frac{1}{\Gamma(2z)}\int^\infty_0d\beta\beta^{2z-1}\frac{\cosh^2\beta}{4\sinh^3\beta}\ ,\nn\\
\widetilde{\zeta}^B_{\text{even,2}}(z)&=&-\frac{1}{\Gamma(2z)}\int^\infty_0d\beta\beta^{2z-1}\frac{1+2\cosh\beta}{4\sinh^3\beta}\ ,\nn\\
\widetilde{\zeta}^B_{{\rm odd}}(z)&=&-\widetilde{\zeta}^B_{{\rm even}\,1}(z)\ .
\eea
With the help of the following identities
\bea
&&\frac{1}{\sinh^3\frac{\beta}{2}}=\frac{2}{\beta^2}\frac{\partial^2}{\partial x^2}\frac{1}{\sinh\frac{\beta x}{2}}~|_{x=1}-\frac{1}{2\sinh\frac{\beta}{2}}\ ,\nn\\
&&4^{-z}\zeta(2z,\frac{a}{2})=\frac{1}{\Gamma(2z)}\int^\infty_0d\beta\beta^{2z-1}\frac{e^{-a\beta}}{1-e^{-2\beta}}\ ,
\eea
where $\zeta(a,b)$ is the Hurwitz zeta function, we finally obtain
\bea
\widetilde{\zeta}^B_{{\rm even}\,1}(z)&=&4^{-(2+z)}\Big[3\zeta(2z,-\ft{1}{2})
+4\zeta(2z-2,-\ft{1}{2})+8\zeta(2z-1,-\ft{1}{2})
\nn\\
&&\qquad\quad+(4^z-1)\zeta(2z)+3(4^z-4)\zeta(2z-2)-4(4^z-2)\zeta(2z-1)\Big]\ ,
\nn\\
\widetilde{\zeta}^B_{{\rm even}\,2}(z)&=&4^{-(1+z)}\Big[-4\zeta(2z-2,0)
-4\zeta(2z-1,0)+(4^z-1)\zeta(2z)\nn\\
&&\qquad\quad-4^z\zeta(2z-2)+4\zeta(2z-1)\Big]\ .
\eea
By using the relation between $F^{(1)}$ and spectral zeta function, one arrives at the results
\bea
F^{(1)}_{{\rm even}\,1} &=& \frac1{16}\left(2\log2-\frac{3\zeta(3)}{\pi^2}\right)\ ,
\qquad
F^{(1)}_{{\rm even}\,2} = \frac1{16}\left(2\log2-\frac{5\zeta(3)}{\pi^2}\right)\ ,
\nn\w2
F^{(1)}_{\rm odd}&=&-F^{(1)}_{{\rm even}\,1}\ .
\label{freo}
\eea
Note that the potential logarithmic divergences in $F^{(1)}_{{\rm even}\,1}$ and $F^{(1)}_{{\rm even}\,2}$ have canceled out, and the above finite results are from $\widetilde{\zeta}^{B\prime}(0)$ terms, in agreement with \cite{Giombi:2013fka}. Furthermore, these results can be used as building blocks for the computation of the free energies of the Konstein-Vasiliev models we are interested in, thanks to the observation that for all those models discussed in Section 2, it is always the case that
\be
n_2 = n_S+n_P\ ,
\ee
where we recall that $n_2$ is number of copies of even fields with $s=2,4,\ldots\infty$, $n_S$ is the number of scalars and $n_P$ is the number of pseudoscalars.

\subsection{Fermions}

We now compute the one loop free energy of all fermionic HS fields. The spectral zeta function of a spin-$s$ fermionic fields is given by
\be
\zeta^F_{(\Delta, s)}(z)=\frac{1}{\Gamma(z)}\int^\infty_0 d\beta
\Big[ \mu(z,\beta)+\nu(z,\beta)\frac{\partial^2}{\partial\alpha^2}\Big]\chi_{\Delta,s}(\beta,\alpha)~\Big\vert_{\alpha=0}\ ,
\label{cf2}
\ee
where
\bea
&&\chi_{\Delta,s}(\beta,\alpha)=\frac{e^{-\beta(\Delta-\ft32)}\sin[(s+\ft12)\alpha]}{4\sinh\frac{\beta}{2}\sin\frac{\alpha}{2}(\cosh\beta-\cos\alpha)}\ ,
\nn\\
&&\mu(z,\beta)=\ft{1}{3}\sinh\ft{\beta}{2}\Big[ f_1(z,\beta) \Big(-6+ \sinh^2\ft{\beta}{2}\Big) +4 f_3(z,\beta) \sinh^2\ft{\beta}{2} \Big]\ ,
\nn\\
&&\nu(z,\beta)=-4f_1(z,\beta)\sinh^3\ft{\beta}{2}\ ,
\nn\\
&&f_n(z,\beta)=
    \sqrt{\pi}\int^\infty_0duu^n\coth(\pi u)(\ft{\beta}{2u})^{z-\ft{1}{2}}J_{z-1/2}(u\beta)\ .
\eea
To compute the one loop free energy of all fermionic HS fields, we propose the following alternate spectral zeta function, which is much easier to
use. The physical equivalence between the alternate spectral zeta function \eqref{c5} and the original one \eqref{cf2} is shown in the appendix.
\be
\widetilde{\zeta}^F_{(\Delta, s)}(z)= \frac{1} {\Gamma(2z)} \int^\infty_0 d\beta  \beta^{2z-1}
\Big[\ft14  \sinh\ft{\beta}{2}+\frac{1}{\sinh\frac{\beta}{2}}+\sinh\ft{\beta}{2} \partial^2_\alpha\Big]
\chi_{\Delta,s}(\beta,\alpha)~\Big\vert_{\alpha=0}\ .
\label{c5}
\ee
The sum of characters of all fermionic HS fields is computed as
\eq{
\label{c9}
\chi_{\frac{3}{2},\frac{1}{2}}( \beta,\alpha)+\sum^\infty_{s=3/2} \Big[\chi_{s+1,s}(\beta,\alpha)
-\chi_{s+2,s-1}(\beta,\alpha)\Big]
=\frac{\cos\frac{\alpha}{2}\cosh\frac{\beta}{2}}{(\cos\alpha-\cosh\beta)^2}\ .
}
It is straightforward to check that
\bea
&&\Big[\ft14  \sinh\ft{\beta}{2}+\frac{1}{\sinh\frac{\beta}{2}}+\left(\sinh\ft{\beta}{2}\right) \partial^2_\alpha\Big]\times\nn\\
&&\Big( \chi_{\frac{3}{2},\frac{1}{2}}( \beta,\alpha)+\sum^\infty_{s=3/2} \Big[\chi_{s+1,s}(\beta,\alpha)
-\chi_{s+2,s-1}(\beta,\alpha)\Big]\Big)\Big\vert_{\alpha=0} =0\ ,
\eea
which indicates that the total one loop free energy of fermionic HS fields in fact vanishes.

\subsection{Summary}

For a Konstein-Vasiliev higher theory consisting of $n_S$ real scalars, $n_P$ pseudoscalars, $n_1$ copies of fields with $s=1,3,...,\infty$, $n_2=n_S+n_P$ copies of fields with $s=2,4,...,\infty$ fields and $n_F$ copies of spin $1/2,3/2,...,\infty$ fields, we have
\be
F^{(1)} (n_S,n_P,n_1,n_2,n_F) = \frac{\log2}8(n_S+n_P-n_1)-\frac{\zeta(3)}{16\pi^2}(3n_S+ 5n_P-3n_1)\ ,
\label{totalF}
\ee
where we have used the relation $n_2=n_S+n_P$. The values of $n_S$, $n_P$ and $n_1$ can be read off from \eqref{KS} for various Konstein-Vasiliev models. Substituting them into the equation above, we obtain
\bea
hu(m;n|4)&:&F^{(1)}_{hu}=-\frac{\zeta(3)}{8\pi^2}n^2\ ,
\label{b1}\\[10pt]
ho(m;n|4)&:& F^{(1)}_{ho}=\frac{\log2}8(m + n)-\frac{\zeta(3)}{16\pi^2}(3 m + 4 n + n^2)\ ,
\label{b2}\\[10pt]
husp(m;n|4)&:& F^{(1)}_{husp}=-\frac{\log2}8(m + n) + \frac{\zeta(3)}{16\pi^2}( 3m+4n-n^2)\ .
\label{b3}
\eea
The one loop free energy of $husp(m;n|4)$ model is related to the one of $ho(m;n|4)$ 
model via $m\rightarrow-m,\,n\rightarrow-n$. The ordinary supersymmetric HS models correspond to the cases $m=n=2^{\frac{\cal N}{2}-1}$ for even ${\cal N}$ and $m=n=2^{({\cal N}-1)/2}$ for odd ${\cal N}$.

As for the ordinary supersymmetric HS models with internal symmetries, we recall that their spectra can be obtained by assigning fundamental representations of the  internal symmetry group to the $OSp({\cal N}|4)$ singletons, and working out the their two-fold tensor products. The resulting spectra are provided in Table 5 of \cite{Sezgin:2012ag}. In particular, the number of fermions with $s=\frac12\, {\rm mod}\, 2$ and  $s=\frac32\, {\rm mod}\, 2$ are the same. As a consequence, the contributions of the fermions to the one loop free energy will continue to vanish since in \eqref{c5} we found that fermions with each half integer spin occurring once give vanishing contribution.
Consequently, the bulk free energy becomes the sum of free energies of type A and type B models with the desired internal symmetries. In this case both $\log2$ and $\zeta(3)$ terms will show up in the one loop free energy, and their coefficients involve $n^2_{\text{fund}}$ dependence, where $n_{\text{fund}}$ is the dimension of fundamental representation of the internal symmetry group. This information is sufficient to perform the one loop test by means of comparing the bulk and boundary free energies, as we shall see at the end of next section.

\section{Free energies of free CFT's on $S^3$ and comparison}

The free energies of free scalars and free fermions which are conformally coupled to $S^3$ have been studied in \cite{Klebanov:2011gs}.
A conformally coupled free scalar and a free fermion on $S^3$ are described by the following two actions respectively
\be
S_S=\frac12\int d^3x\sqrt{g}\Big[(\nabla\phi)^2+\frac3{4L^2}\phi^2\Big]\ ,\quad S_D=\frac12\int d^3x\sqrt{g}\psi^{\dagger}({\rm i} \slashed{D}\psi)\ ,
\eea
where $L$ is the radius of the round $S^3$. Free energies of the above two theories are defined as usual
\bea
F_S&=&-\log Z_S=\frac12 \log{\rm det}[\Lambda^{-2}{\cal O}_S]\ ,\quad {\cal O}=-\nabla^2+\frac3{4L^2}\ ,\nn\\
F_D&=&-\log Z_D=-\log{\rm det}[\Lambda^{-1}{\cal O}_D]\ ,\quad {\cal O}={\rm i}\slashed{D}\ .
\eea
Using zeta function, $F_S$ and $F_D$ can be computed straightforwardly and the results are \cite{Klebanov:2011gs}
\be
F_S=\frac1{16}\Big(2\log2-\frac{3\zeta(3)}{\pi^2}\Big)\ ,\quad
F_D=\frac1{8}\Big(2\log2+\frac{3\zeta(3)}{\pi^2}\Big)\ .
\label{fsd}
\ee
Notice that the free energy of a Majorana fermion on $S^3$ is $\ft12F_D$.


A bulk HS theory is conjectured to be dual to a free vector model when the boundary conditions of the bulk fields 
preserve the HS symmetry \cite{Sezgin:2002rt,Klebanov:2002ja}, which is the case here.
Assuming the bulk HS theory possesses an action, its free energy associated with $AdS_4$ should have the form
displayed in \eqref{Fbulk}
%
%
where $G_N$ is the Newton's constant. In cases where the boundary of $AdS_4$ is $S^3$, the bulk free energy 
should be compared with that of a free vector model on $S^3$ order by order in $1/N$ expansion. Hence the 
comparison requires an identification between $G_N$ and $N$. It was suggested by \cite{Giombi:2013fka} that 
in general the relation between $G_N$ and $N$ is of the form given in \eqref{renorm} 
%
%
where $\gamma$ and $\Delta N$ are constants and especially $\Delta N$ should be an integer. The basic fields in the vector model constitute a vector
in the fundamental representation of a classical Lie group, which can be $U(N)$, $O(N)$ or $USp(N)$ in our cases.
The free energy of a free vector model can be computed exactly and be put in the form\footnote{Strictly speaking, 
the bulk HS theory is dual to the $U(N)$, $O(N)$ or $USp(N)$ singlet sector of a free CFT. The partition 
function of a free CFT on $S^3$ is evaluated in the vacuum which is already a singlet state under the corresponding 
symmetry group in each case. Thus, imposing the singlet constraint should not affect the free energy. }
\be
F_{\small{ \rm CFT}}=NF^{(0)}_{\small{ \rm CFT}}\ ,
\ee
where we use $F^{(0)}_{\small{ \rm CFT}}$ to denote the contribution of a single component in the vector.
For $F_{\rm bulk}$ to match with $F_{\small{ \rm CFT}}$, it is clear that the bulk free energy at each order in $G_N$ expansion should all be
proportional to $F^{(0)}_{\small{ \rm CFT}}$.

Various one loop tests of HS holography have been carried out in the literature \cite{Giombi:2013fka,Giombi:2014iua}.
For instance, the non-minimal type A model is conjectured to be dual to the $U(N)$ singlet sector of $N$ complex scalars. 
When HS symmetry is preserved by the boundary condition, $F^{(1)}_{\rm bulk}$ was found to be 0, 
indicating that $G^{-1}_N$ is identified with $N$ at one loop order. For minimal A model, the conjectured dual CFT is 
the $O(N)$ singlet sector of $N$ real scalars. In this case, $F^{(1)}_{\rm bulk}$ is equal to $F_S$, the free energy of a real free 
scalar \eqref{fsd}. Thus, matching the bulk and boundary free energies at one loop order requires $G^{-1}_N$ being identified 
with $N-1$. The $husp(2;0|4)$ Vasiliev theory is conjectured to be dual to the $USp(N)$ singlet sector of $N$ complex 
scalars and $F^{(1)}_{\rm bulk}$ is equal to $-F_S$. Therefore, for $husp(2;0|4)$ higher spin theory, $G^{-1}_N$ is 
identified with $N+1$ at one loop order.

In this section, we consider the cases in which the bulk HS symmetry 
is preserved by the boundary condition, thus the CFT duals are certain singlet sectors of free CFTs composed by free 
scalars and free fermions. For the $hu(m;n|4)$ theory, the dual CFT consists of $Nm$ complex free scalars 
$\phi^{ia}$, $i=1,2,...N$, $a=1,2,...m$ and $Nn$ Dirac fermions $\psi^{ir}$, $r=1,2,...n$. The $m^2$ $\Delta=1$ scalars and $n^2$ $\Delta=2$ pseudoscalars correspond to the operators
\be
\bar{\phi}_{ia}\phi^{ib}\ ,\qquad \bar{\psi}_{ia}\psi^{ib}\ .
\ee
Free energy of this theory is given by
\be
F_{\small{ \rm CFT}}=NF^{(0)}_{\small{ \rm CFT}}\ ,\quad F^{(0)}_{\small{ \rm CFT}}=2mF_S+nF_D\ ,
\label{UnF}
\ee
where $F_S$ and $F_D$ are given in \eqref{fsd}.

For the $ho(m;n|4)$ theory, the dual CFT consists of $Nm$ real free scalars $\phi^{ia}$, $i=1,2,...N$, $a=1,2,...m$ and $Nn$ majorana fermions $\psi^{ir}$, $r=1,2,...n$. The $m^2$ $\Delta=1$ scalar fields and $n^2$ $\Delta=2$ pseudoscalars correspond to the operators
\be
\phi^{ia}\phi^{jb}\delta_{ij}\ , \qquad \bar{\psi}^{ia}\psi^{jb}\delta_{ij}\ .
\ee
The free energy is given by
\be
F_{\small{ \rm CFT}}=NF^{(0)}_{\small{ \rm CFT}}\ ,\quad F^{(0)}_{\small{ \rm CFT}}=m F_S+\ft12n F_D\ .
\label{OnF}
\ee
For the $husp(m;n|4)$ theory, the dual CFT consists of $Nm$ complex free scalars $\phi^{ia}$, $i=1,2,...N$, $a=1,2,...m$ 
and $Nn$ Dirac fermions $\psi^{ir}$, $r=1,2,...n$, subject to the symplectic reality condition.  The $m^2$ $\Delta=1$ scalar 
fields and $n^2$ $\Delta=2$ pseudoscalars correspond to the operators
\be
\phi^{ia}\phi^{jb}\Omega_{ij}\ ,\qquad \bar{\psi}^{ia}\psi^{jb}\Omega_{ij}\ ,
\ee
where $\Omega_{ij}$ is the $USp(N)$ invariant tensor. Free energy of this theory is given by
\be
F_{\small{ \rm CFT}}=NF^{(0)}_{\small{ \rm CFT}}\ ,\quad F^{(0)}_{\small{ \rm CFT}}=m F_S+\ft12n F_D\ .
\label{UspF}
\ee

Since supersymmetric HS theories can be mapped to special cases of Konstein-Vasiliev models, we will not give 
separate discussions on them. 

As discussed before, duality between the bulk HS theory and boundary free CFT may be achieved only if 
$F^{(1)}_{\rm bulk}$ is proportional to $F^{(0)}_{\rm CFT}$. Using \eqref{totalF}, \eqref{fsd}, \eqref{UnF}, \eqref{OnF} and \eqref{UspF}, 
we find that this requirement amounts to
\be
(m+n)(3n_S+5n_P-3n_1)=3(m-n)(n_S+n_P-n_1)\ ,
\label{f1f0}
\ee
obtained by setting the ratios of $\log 2$ and $\xi(3)$ dependent terms equal to each other.  Taking the values 
of $n_S$, $n_P$ and $n_1$  from \eqref{KS}, these ratios for the bulk sides can be read off from \eqref{b1}, 
\eqref{b2} and \eqref{b3} in terms of $m$ and $n$. One can show that for all three Konstein-Vasiliev models, the only solution to the equation above  is given by $n=0$, which implies bosonic type A models. In this case the $\log 2$ and $\zeta(3)$ dependent terms arise in the same ratio as of a single real scalar field, and we have the result 
\be
F^{(1)}_{hu(m;0|4)}=0\ ,\qquad
F^{(1)}_{ho(m;0|4)}=mF_S\ ,\qquad F^{(1)}_{ho(m;0|4)}=-mF_S\ .
\ee
Therefore, assuming that $F^{(0)}_{\rm bulk}=F^{(0)}_{\rm CFT}$, the bulk and 
boundary free energies match with each other provided that
\bea
&& hu(m;0|4):\, G^{-1}_N\rightarrow N\ ,
\nn\\
&& ho(m;0|4):\, G^{-1}_N\rightarrow N-1\ ,
\nn\\
&& husp(m;0|4):\,G^{-1}_N\rightarrow N+1\ .
\eea
The holographic dictionaries relating $G_N$ to $N$ in various HS models 
have been put forward in \cite{Giombi:2013fka} via testing the holography of $hu(1;0|4)$, 
$ho(1;0|4)$ and $husp(2;0|4)$ models  at one loop level. Here, we have extended the 
validity of these holographic mappings to $hu(m;0|4)$, $ho(m;0|4)$ and  $husp(m;0|4)$ Konstein-Vasiliev models. We see that the inclusion of infinite tower of bulk fermions 
does not cure the problem with the mismatch of the free energies in the type B model, 
which corresponds to the case in which $m=0$ and $n\ne 0$, and its conjectured dual. 

Finally, we consider the ordinary supersymmetric models with internal symmetry discussed earlier, whose spectra are given in Table 5 of \cite{Sezgin:2012ag}. In Section 3 we found that the contributions of the bulk fermions give vanishing contributions to free energy and consequently the bulk free energy becomes the sum of free energies of type A and type B models with the desired internal symmetries. Furthermore, we noted that the coefficient of the $\zeta(3)$ dependent contribution to the free energy will have $n_{\rm fund}^2$ dependence, where $n_{\rm fund}$ is the dimension of the fundamental representation of the internal symmetry group. On the other hand it is easy to show that the $\zeta(3)$ dependent terms on the CFT side vanish.  Therefore, we conclude the problem of free energy mismatch will persist in ordinary supersymmetric HS theories with internal symmetry.


\section{One loop free energies of supersymmetric higher spin theories in $AdS_4$ with $S^1_{\beta}\times S^2$ boundary}


In thermal $AdS_4$, the one loop free energy of the bulk theory takes the form \cite{Giombi:2014yra}
\be
F^{(1)}_{\rm bulk}=F(\beta)_{\rm bulk}+\beta E_{c\,{\rm bulk}}+a_{\rm bulk}\log\Lambda\ ,
\label{tpf1}
\ee
where $\beta$ is the period of the imaginary time, $F(\beta)_{\rm bulk}$ is the 
thermal free energy which can be computed by taking the log of the thermal partition 
function as $F(\beta)_{\rm bulk}\equiv\beta^{-1}\log Z_{\rm bulk}$ with 
$Z_{\rm bulk}\equiv{\rm tr} \,e^{-\beta H_{\rm bulk}}$, and $a_{\rm bulk}$ is 
the anomaly coefficient related to the Seeley coefficient. The trace denotes the sum over all HS particle states. $a_{\rm bulk}$ is proportional to the integral of local curvature invariants, and should be the same for $AdS_4$ with $S^3$ boundary and for 
the thermal $AdS_4$. Thus, after summing over spins the total $a_{\rm bulk}$ 
should vanish as shown in previous sections.
$E_{c\,{\rm bulk}}$ 
is the one loop contribution to the Casimir energy which can be extracted from the 
thermal free energy in a standard way (cf. \eqref{bulkcasimir}, \eqref{zetacasimir}). 

The free energy of the $U(N),\,O(N)$ or $USp(N)$ singlet sector 
of a free vectorial CFT on $S^1_{\beta}\times S^2$ takes similar form
\be
F_{\rm CFT}=F^{\rm singlet}(\beta)_{\rm CFT}+\beta E_{c\,{\rm CFT}}
+a_{\rm CFT}\log\Lambda\,,
\ee
in which $F(\beta)_{\rm CFT}$ is the free energy of the subsector in Hilbert 
space consisting of only the states that are invariant under the required symmetry 
group. The Casimir energy $E_{c\,{\rm CFT}}$ is given by $NE_0$, where $E_0$ is the Casimir energy of a single 
conformally invariant free field on $S^1_{\beta}\times S^2$. The anomaly coefficient 
$a_{\rm CFT}$ vanishes on $S^1_{\beta}\times S^2$, which is conformally flat and has 
vanishing Euler number. Therefore, there are no logarithmic divergent terms on both 
the bulk and the boundary sides. There remains comparison of the thermal part of the free 
energies and the Casimir energies on both sides. The thermal part of the free energies are expected to match since, by definition, the bulk and boundary thermal partition 
functions which give rise to the corresponding thermal free energies are both equal to 
the character of the HS algebra associated with the spectrum of the HS theory. 
The comparison between the bulk and boundary Casimir energies, however, 
is not straightforward, since different from $E_{c\,{\rm bulk}}$, the Casimir energy 
on the CFT side is not directly related to the thermal free energy of the singlet 
sector through \eqref{bulkcasimir}. Holographic matching of the free energies at 
${\cal O}(N^0)$ demands that $E_{c\,{\rm bulk}}$ is an integer times the Casimir 
energy of a single conformally invariant free field on $S^1_{\beta}\times S^2$.

In this section, we first study the one loop free energy of Konstein-Vasiliev theory 
in thermal $AdS_4$ with $S^1_{\beta}\times S^2$ boundary. We then compare the bulk result with the free energy of the corresponding dual CFT at ${\cal O}(N^0)$. Recall that there exist generalizations of $d>4$ Vasiliev theory which are dual to the $U(N)$ or $O(N)$ singlet sector of free scalars or fermions \cite{Vasiliev:2003ev}. Free energy of this type of HS theory in thermal 
$AdS_{d}$ has been calculated in \cite{Giombi:2014yra} and compared with $O(N^0)$ term in the free energy of the large $N$ $U(N)$ or $O(N)$ vectorial 
free CFT. It was found that the matching of free energy implies shifts in the 
relation between $G^{-1}_N$ and $N$ at leading order by an integer. 

Different from \cite{Giombi:2014yra} where the bulk theories are purely bosonic, in our case the bulk theory includes also fermionic HS fields. Accordingly, 
the dual CFT consists of both scalars and fermions. In particular, the fermionic 
HS fields are dual to the bilinear conserved currents built out of both scalars and 
fermions. State operator correspondence then implies the existence of scalar-fermion 
mixed states in the Hilbert space that are singlet under the required symmetry group. 
These scalar-fermion mixed states contribute to the thermal free energy of the 
singlet sector nontrivially, which means that the $F^{\rm singlet}(\beta)$ for a 
CFT involving both scalars and fermions cannot be obtained by a simple sum of the $F^{\rm singlet}(\beta)$'s of a pure-scalar CFT and of a pure-fermion CFT.

Below we start with the computation of the free energies in Konstein-Vasiliev models, 
which include supersymmetric HS theories as special cases. The story is far more 
elaborate in higher dimensions.  In particular, we refer the reader to 
\cite{Beccaria:2014xda,Beccaria:2014zma}  and \cite{Bae:2016hfy} for the case of 5D, and 
\cite{Beccaria:2014qea} for the case of 7D.

\subsection{ The bulk side}

As stated earlier, the one loop free energy of a massless field in thermal $AdS_4$ 
has the structure
displayed in \eqref{tpf1} with the vanishing log divergence. $F(\beta)$ can be obtained from the grand canonical partition function as
\bea
\label{a2}
\text{For bosons:}~~F(\beta)_{\rm bulk}
&=& -\sum^\infty_{m=1}\frac{1}{m}\mathcal{Z}(m\beta)\ ,
\\
\text{For fermions:}~~F(\beta)_{\rm bulk}
&=&\sum^\infty_{m=1}\frac{(-1)^m}{m}\mathcal{Z}(m\beta)\ .
\eea
Here $\mathcal{Z}(\beta)$ is the one-particle canonical partition function.
The Casimir energy $E_{c\,{\rm bulk}}$ can be obtained from the energy $\zeta$-function as
\eq{
E_{c\,{\rm bulk}} =\pm\frac{1}{2}\zeta_E(-1)\ ,
\label{bulkcasimir}
}
where $\pm$ correspond to bosonic and fermionic cases respectively. 
The energy $\zeta$-function is related to the one-particle partition function by a Mellin transform
\eq{
\zeta_E(z)=\frac{1}{\Gamma(z)}\int^\infty_0d\beta\beta^{z-1}\mathcal{Z}(\beta)\ .
\label{zetacasimir}
}
In $d=4$, the thermal one-particle 
partition function for a scalar field is given by
\eq{
\mathcal{Z}^{(\Delta)}_0=\frac{q^\Delta}{(1-q)^3}\quad \Delta>\frac12\ ,
}
where $\Delta$ is the $AdS$ energy and $q=e^{-\beta}$ \cite{Gibbons:2006ij}.
Thermal one-particle partition function for $s\ge\ft12$ massless field takes the form
\eq{
\mathcal{Z}_s(\beta)=\frac{q^{s+1}}{(1-q)^3}\big[2s+1-(2s-1)q\big]\ .
}
From the results derived in \cite{Giombi:2014yra}, we deduce the useful formulae\footnote{In the rest of this subsection the thermal free energies 
and partition functions refer to those of the bulk theory.}
\bea
F^{(1)}_{{\rm even}\,1}&=&F(\beta)_{{\rm even}\,1}
=-\sum^\infty_{m=1}\frac{1}{m}\mathcal{Z}_{{\rm even}\,1}(m\beta)\ ,
\nn\\
\mathcal{Z}_{{\rm even}\,1}(\beta)&=&\ft12\frac{q(1+q)^2}{(1-q)^4}
+\ft12\frac{q(1+q^2)}{(1-q^2)^2}=\ft12[\widetilde{\mathcal{Z}}_0(\beta)]^2
+\ft12\widetilde{\mathcal{Z}}_0(2\beta)\ ,
\nn\\
F^{(1)}_{{\rm even}\,2}&=&F(\beta)_{{\rm even}\,2}
=-\sum^\infty_{m=1}\frac{1}{m}\mathcal{Z}_{{\rm even}\,2}(m\beta)\ ,
\nn\\
\mathcal{Z}_{{\rm even}\,2}(\beta)&=&\frac{2q^2}{(1-q)^4}
-\frac{q^2}{(1-q^2)^2}=\ft12[\widetilde{\mathcal{Z}}_{\frac12}(\beta)]^2
-\ft12\widetilde{\mathcal{Z}}_{\frac12}(2\beta)\ ,
\nn\\
F^{(1)}_{{\rm odd}\,1}&=&F(\beta)_{{\rm odd}}
=-\sum^\infty_{m=1}\frac{1}{m}\mathcal{Z}_{\rm odd}(m\beta)\ ,
\nn\\
\mathcal{Z}_{\rm odd}(\beta)&=&\ft12\frac{q(1+q)^2}{(1-q)^4}
-\ft12\frac{q(1+q^2)}{(1-q^2)^2}=\ft12[\widetilde{\mathcal{Z}}_0(\beta)]^2
-\ft12\widetilde{\mathcal{Z}}_0(2\beta)\ ,
\label{freo2}
\eea
where for later convenience we express the results in terms of the characters $\widetilde{\mathcal{Z}}_0(\beta)$ and $\widetilde{\mathcal{Z}}_{\ft12}(\beta)$ of the conformally coupled free scalar and the free real fermion which realize the spin-0 and spin-$\ft12$ singleton representations of the $SO(3,2)$, respectively
\be
\widetilde{\mathcal{Z}}_0(\beta)=\frac{q^\ft12 (1+q)}{(1-q)^2}\ ,\qquad
\widetilde{\mathcal{Z}}_{\ft12}(\beta)=\frac{2q}{(1-q)^2}\ .
\ee
By using (\ref{bulkcasimir}) and (\ref{zetacasimir}), one can show that $\mathcal{Z}_{{\rm even}\,1}(\beta)$, $\mathcal{Z}_{{\rm even}\,2}(\beta)$ and $\mathcal{Z}_{\rm odd}(\beta)$ all lead to vanishing Casimir energy \cite{Giombi:2014yra}. Therefore we simply dropped $E_c$ term in (\ref{freo2}). Also
one should note that
\be
\ft12[\widetilde{\mathcal{Z}}_{\frac12}(\beta)]^2+\ft12\widetilde{\mathcal{Z}}_{\frac12}(2\beta)
=\ft12[\widetilde{\mathcal{Z}}_0(\beta)]^2-\ft12\widetilde{\mathcal{Z}}_0(2\beta)\ .
\ee
For all the fermionic fields, we find that the total one-particle canonical partition 
function is given by
\be
\mathcal{Z}^F(\beta)=\sum^\infty_{s=\frac{1}{2}}\frac{q^{s+1}}{(1-q)^3}
\Big[2s+1-(2s-1)q\Big]=\frac{2q^{\frac{3}{2}}(1+q)}{(1-q)^4}
=\widetilde{\mathcal{Z}}_0(\beta)\widetilde{\mathcal{Z}}_{\frac12}(\beta)\ .
\eea
Using the total one-particle canonical partition function, we can construct the energy $\zeta$-function for fermions
\bea
\zeta_E^F(z)&=&\frac{1}{\Gamma(z)}\int^\infty_0d\beta\beta^{z-1}
\frac{2e^{-\frac{3}{2}\beta}(1+e^{-\beta})}{(1-e^{-\beta})^4}
\nn\\
&=&2\sum^\infty_{n=1}{n+2\choose3}[(n+\ft{1}{2})^{-z}+(n+\ft{3}{2})^{-z}]
\nn\\
&=&\ft{1}{8}\zeta(z,\ft{5}{2})-\ft{1}{12}\zeta(z-1,\ft{5}{2})
-\ft{1}{2}\zeta(z-2,\ft{5}{2})+\ft{1}{3}\zeta(z-3,\ft{5}{2})
\nn\\
&&-\ft{1}{8}\zeta(z,\ft{3}{2})-\ft{1}{12}\zeta(z-1,\ft{3}{2})
+\ft{1}{2}\zeta(z-2,\ft{3}{2})+\ft{1}{3}\zeta(z-3,\ft{3}{2})\ .
\eea
This vanishes at $z=-1$. Therefore, the total Casimir energy for fermionic 
HS fields vanishes in thermal $AdS_4$ as well, and the correspoding one loop free energy is simply
\be
F^{(1)\, F}=F(\beta)^F_{\rm bulk}
=\sum^\infty_{m=1}\frac{(-1)^m}{m}\mathcal{Z}^F(m\beta)\ .
\ee
Summarizing the results above and using the spectra given in \eqref{KS}, we
find that the one loop free energies for generic Konstein-Vasiliev HS theories are given by
\bea
hu(m;n|4)&:& F^{(1)}_{hu}=-\sum^\infty_{k=1}\frac1{k}\Big[m\,
\widetilde{\mathcal{Z}}_0(k\beta)+n\,(-)^{k+1}\,\widetilde{\mathcal{Z}}_{\frac12}(k\beta)\Big]^2\ ,
\label{fhu2}
\\[10pt]
ho(m;n|4)&:& F^{(1)}_{ho}=-\sum^\infty_{k=1}\frac1{2k}
\Big(\Big[m\,\widetilde{\mathcal{Z}}_0(k\beta)+n\,(-)^{k+1}\,
\widetilde{\mathcal{Z}}_{\frac12}(k\beta)\Big]^2\nn\\
&&\qquad \qquad +m\,\widetilde{\mathcal{Z}}_0(2k\beta)-n\,
\widetilde{\mathcal{Z}}_{\frac12}(2k\beta)\Big)\ ,
\label{fho2}
\\[10pt]
husp(m;n|4)&:& F^{(1)}_{husp}=-\sum^\infty_{k=1}\frac1{2k}\Big(\Big[m\,
\widetilde{\mathcal{Z}}_0(k\beta)+n\,(-)^{k+1}\,\widetilde{\mathcal{Z}}_{\frac12}(k\beta)\Big]^2
\nn\\
&&\qquad \qquad -m\,\widetilde{\mathcal{Z}}_0(2k\beta)+n\,
\widetilde{\mathcal{Z}}_{\frac12}(2k\beta)\Big)\ .
\label{fhusp2}
\eea
The free energy of $husp(m;n|4)$ theory can be obtained from that of the $ho(m;n|4)$ theory by $m\rightarrow-m\,,n\rightarrow-n$.

\subsection{The CFT side and comparison}

In this section, we calculate the partition function of the singlet sector of free CFTs on $S^1_{\beta}\times S^2$. 
We closely follow the technique developed in \cite{Aharony:2003sx,Schnitzer:2004qt}. The partition function
of a CFT on $S^1_{\beta}\times S^2$ is equal to the thermal partition function 
due to the vanishing of Casimir energy and logarithmic divergence. Therefore, we have
\be
Z(\beta)=\sum_{i\in {\rm physical\,states}} q^{E_i}\ ,\quad q=e^{-\beta}\ ,
\ee
where the ${\rm physical\,states}$ are restricted to be the singlet states of 
$U(N)$, $O(N)$ or $USp(N)$ for our purpose. We have also used the fact that there is no non-trivial chemical potential in the system. The thermal partition
functions of the $U(N)$ and $O(N)$ singlet sectors of free scalar and free fermion theories have been studied in \cite{Shenker:2011zf,Giombi:2014yra}. We generalize their results to the cases with both scalars and fermions. We first consider 
the $U(N)$ singlet sector of a free CFT with $Nm$ complex free scalars and $Nn$ Dirac fermions. As shown in \cite{Shenker:2011zf,Giombi:2014yra}, the thermal 
partition function can be expressed as a path integral localized on the eigenvalues of $U(N)$ matrix
\bea
&&Z_{U(N)}(\beta)=e^{-F(\beta)_{U(N)}}=\int\prod_{i=1}^N\,d\alpha_i e^{-S(\alpha_1,...\alpha_N)}\ ,
\nn\\
&&S(\alpha_1,...\alpha_N)=-\ft12\sum^{N}_{i\neq j=1}\log\sin^2\frac{\alpha_i-\alpha_j}2+2\sum^N_{i=1}f_{\beta}(\alpha_i)\ ,
\nn\\
&&f_{\beta}(\alpha)=\sum^N_{k=1}c_k(\beta)\cos(k\alpha)\ ,\quad c_k(\beta)
=-\frac1k\Big[m\,\widetilde{\mathcal{Z}}_0(k\beta)+n\,(-)^{k+1}\widetilde{\mathcal{Z}}_{\frac12}(k\beta)\Big]\ ,
\label{unea}
\eea
where the matter contents affect the effective action through $c_k(\beta)$. 
In the large $N$ limit, the integral over $\alpha_i$ can be replaced by the 
path integral over the eigenvalue density $\rho(\alpha)$, $\alpha\in (-\pi,\pi)$. 
$\rho(\alpha)$ satisfies the standard normalization
\be
\int^{\pi}_{-\pi}d\alpha\rho(\alpha)=1\ .
\ee
The effective action in terms of $\rho(\alpha)$ takes the form
\bea
S(\rho)&=&N^2\int d\alpha d\alpha' K(\alpha-\alpha')\rho(\alpha)\rho(\alpha')+2N\int d\alpha\rho(\alpha)f_{\beta}(\alpha)\ ,\nn\\
 K(\alpha-\alpha')&=&-\ft12\log(2-2\cos\alpha)\ ,\qquad f_{\beta}(\alpha)=\sum^N_{k=1}c_k(\beta)\cos(k\alpha)\ .
\eea
Integrating out $\rho$, one obtains
\be
F(\beta)_{U(N)}=-\sum^{\infty}_{k=1}k[c_k(\beta)]^2 
=-\sum^{\infty}_{k=1}\frac1k\Big[m\,\widetilde{\mathcal{Z}}_0(k\beta)+n\,(-)^{k+1}\widetilde{\mathcal{Z}}_{\frac12}(k\beta)\Big]^2\ ,
\ee
which coincides with one loop free energy for $hu(m;n|4)$ higher spin theory \eqref{fhu2}.
Next, we study the $O(N)$ singlet sector of a free CFT with $Nm$ real free scalars and $Nn$ Majorana fermions. 
This is a generalization of the results in \cite{Giombi:2014yra}, where the free CFT consists of only scalars or fermions. 
It is suggested in \cite{Giombi:2014yra} that, one can choose $N$ to be even, 
namely $N$=2N for simplicity in the large $N$. The difference between even $N$ 
and odd $N$ cases is at the next order in $1/N$ expansion. Free energy of the 
$O(2{\rm N})$ singlet sector of a free CFT with $Nm$ real free scalars and 
$Nn$ Majorana fermions can again be written as a path integral over the eigenvalues 
of $O(N)$ matrix. The effective potential of the $O(N)$ singlet sector is given by \cite{Giombi:2014yra}
\be
S(\alpha_1,...\alpha_{\rm N})=-\ft12\sum^{{\rm N}}_{i\neq j=1}\log\sin^2\frac{\alpha_i-\alpha_j}2
-\ft12\sum^{{\rm N}}_{i\neq j=1}\log\sin^2\frac{\alpha_i+\alpha_j}2+2\sum^{\rm N}_{i=1}f_{\beta}(\alpha_i)\ ,
\ee
where $f_{\beta}$ is the same as the one in \eqref{unea}. The effective potential 
for the $O(N)$ singlet sector differs from that of the $U(N)$ by the $\log\sin^2\alpha$ 
terms which come from the Van der Monde determinant or the Haar measure. In the large $N$ limit, the path integral over $\alpha_i$ can again be recast into an integral over the eigenvalue density $\rho(\alpha)$. After integrating out $\rho$, one obtains
\be
F(\beta)_{O(N)}&=&-\sum^{\infty}_{k=1}\frac{k}2\Big([c_k(\beta)]^2
-\ft2k\,c_{2k}(\beta)\Big)
\\
&=&-\sum^{\infty}_{k=1}\frac1{2k}\Big(\Big[m\,\widetilde{\mathcal{Z}}_0(k\beta)+n\,(-)^{k+1}\,
\widetilde{\mathcal{Z}}_{\frac12}(k\beta)\Big]^2
+m\,\widetilde{\mathcal{Z}}_0(2k\beta)-n\,\widetilde{\mathcal{Z}}_{\frac12}(2k\beta)\Big)\ ,
\nn
\ee
which matches the one loop free energy of $ho(m;n|4)$ HS theory in \eqref{fho2}. 
In the last case, we consider the $USp(N)$ singlet sector of a free CFT with $Nm$ 
complex free scalars $\phi^{ia}$, $i=1,2,...N$, $a=1,2,...m$ and $Nn$ Dirac fermions 
subject to the symplectic real condition. Since $N$ is even in this case, we denote $N$ 
by $2{\rm N}$. The effective potential of the $USp(N)$ singlet sector takes the form
\be
S(\alpha_1,...\alpha_{\rm N})&=&-\ft12\sum^{{\rm N}}_{i\neq j=1}\log\sin^2\frac{\alpha_i-\alpha_j}2
-\ft12\sum^{{\rm N}}_{i, j=1}\log\sin^2\frac{\alpha_i+\alpha_j}2
\nn\\
&&-\ft12\sum^{{\rm N}}_{i=1}\log\sin^2\alpha_i+2\sum^{\rm N}_{i=1}f_{\beta}(\alpha_i)\ .
\ee
In the large $N$ limit, the path integral over $\alpha_i$ can be evaluated by 
using the same technique as before. The free energy of the $USp(N)$ singlet sector 
of a free CFT is obtained as
\be
F(\beta)_{USp(N)}&=&-\sum^{\infty}_{k=1}\frac{k}2\Big([c_k(\beta)]^2+\ft2k\,c_{2k}(\beta)\Big)
\\
&=&-\sum^{\infty}_{k=1}\frac1{2k}\Big(\Big[m\,\widetilde{\mathcal{Z}}_0(k\beta)+n\,(-)^{k+1}\,
\widetilde{\mathcal{Z}}_{\frac12}(k\beta)\Big]^2
-m\,\widetilde{\mathcal{Z}}_0(2k\beta)+n\,\widetilde{\mathcal{Z}}_{\frac12}(2k\beta)\Big)\ ,
\nn
\ee
which matches one loop free energy of $husp(m;n|4)$ HS theory in \eqref{fhusp2}.

\section{ Mixed boundary conditions in bulk and interacting ${\cal N}=1$ SCFT}

In $\cN=1$ HS theory, the $OSp(1|4)$ invariant boundary conditions are given 
in \cite{Sezgin:2003pt}\footnote{Here we correct a sign error in the result given by \cite{Sezgin:2003pt}.}. To describe this, we write the boundary behavior $(\rho\rightarrow0)$ of the complex scalar $\phi=A+{\rm i} B$ as
\eq{
A=\rho \alpha_+ +\rho^2\beta_+\ , \quad B=\rho \alpha_- +\rho^2\beta_-\ ,
}
and define the 3d, $\cN=1$ superfields
\eq{
\Phi_-=\alpha_-+{\rm i} \bar{\theta}\eta_--\frac{\bar{\theta}\theta}{2\rm i}\beta_+\ ,\quad
\Phi_+=\alpha_++{\rm i} \bar{\theta}\eta_++\frac{\bar{\theta}\theta}{2\rm i}\beta_-\ .
}
The boundary conditions preserving $OSp(1|4)$ take the form
\eq{
\Phi_-=\lambda \Phi_+\ ,
\label{bc1}
}
where $ \lambda $ is an arbitrary real number. In terms of the new scalar fields we have
\eq{
A'=\sin\vartheta  A-  \cos\vartheta B\ ,\quad  B'=\cos\vartheta  A+ \sin\vartheta B\ ,
}
where $ \tan \vartheta = \lambda$, and the boundary condition \eqref{bc1} is equivalent to
\eq{
\alpha'_+=0\ ,\quad \beta'_-=0\ .
\label{bc2}
}
The linearized bulk scalar field equations would remain the same form under the $SO(2)$ rotation, thus the newly defined scalar fields $A'$ and $B'$ possess the same Feffer-Graham expansion as the original scalar fields $A$ and $B$. The boundary condition \eqref{bc2} implies that near the boundary
\eq{
A'=\rho^2\beta_+'\ , \quad  B'=\rho \alpha_-'\ .
}
Therefore, in computing the one loop free energy, $A'$ should have $\Delta=2$, while  $B'$ should have $\Delta=1$, which does not affect the $\mathcal{N}=1$ HS spectrum and the corresponding one loop calculation. On the CFT side, the boundary condition 
\eqref{bc1} implies the $\cN=1$ free CFT being deformed by a supersymmetric double-trace term
\eq{
\Delta S=\frac{\lambda}2 \int d^3xd^2\theta {\cal O}^2\ ,
}
where ${\cal O}$ is given by
\eq{
{\cal O}=\frac1{\sqrt N} W^2\ , \quad W=\varphi +{\rm i}\bar{\theta}\psi+\frac{\bar{\theta}\theta}{2\rm i}f\ .
}
We compute the difference between the free energy of the 
deformed CFT and that of the free CFT, following the procedure adopted in \cite{Klebanov:2011gs,  Gubser:2002vv}.  
Denoting the partition function of the free CFT by $Z_0$, we calculate
\eq{
\Delta F= -\log \frac{Z}{Z_0}\ .
}
Using the Hubbard-Stratonovich transformation, we have
\eq{
\frac{Z}{Z_0}=\frac1{\int D\Sigma {\rm exp}(\frac1{2\lambda}\int dz' \Sigma^2)}
\int D\Sigma \Big\langle {\rm exp}\Big[ \int dz\Big (\frac1{2\lambda}\Sigma^2+\Sigma {\cal O}\Big)\Big]\Big\rangle_0\ ,
\label{deltaF}
}
where $\Sigma$ is an auxiliary superfield and $z$ denotes the supercoordinate. In the large $N$ limit, the higher point functions of
$\cal O$ are suppressed. This allows us to write
\eq{
 \Big\langle {\rm exp}\Big[\int dz\Sigma {\cal O} \Big]\Big\rangle_0
={\rm exp}\Big[\frac12\Big\langle \Big(  \int dz \Sigma  {\cal O}\Big)^2\Big\rangle_0+o(1/N)\Big]~.
}
Note that $\Sigma$ and $\cal O$ are single-trace operators of $\mathcal{N}=1$ superfields, say $M$ and $W$ 
respectively, each with component fields $A^i,\lambda^i,B^i$ and $\phi^i,\psi^i,f^i$, where $B$ and $f$ are auxiliary fields, 
and the index $i$ stands for the representation of $O(N)$. The component fields obey the following superconformal transformations
\al{
&\delta A=\frac{1}{4}\xi\lambda~~~~~~~~~~~~~~~~~~~~~~~\delta\phi=\frac{1}{4}\xi\psi\\
&\delta\lambda =\slashed{\partial}A\xi-\frac{1}{4}B\xi+A\eta~~~~~~\delta\psi =\slashed{\partial}\phi\xi-\frac{1}{4}f\xi+\phi\eta\\
&\delta B=-\xi\slashed{\nabla}\lambda~~~~~~~~~~~~~~~~~~~~\delta f=-\xi\slashed{\nabla}\psi
}
where $\xi$ and $\eta$ are spinors satisfying the conformal Killing spinor equation $\nabla_\mu\xi=\gamma_\mu\eta$.

Integrating out the spinor coordinates $\theta$ and $\bar{\theta}$, we obtain
\eq{
\spl{
\int dz\frac{1}{2\lambda}\Sigma^2=&\frac{1}{\lambda}\int dx^3\sqrt{g}(B^iA^iA^jA^j
+\frac{1}{2}\lambda^i\lambda^iA^jA^j+\lambda^i\lambda^jA^iA^j)\\
=&\frac{1}{\lambda}\int dx^3\sqrt{g}(\Sigma_2\Sigma_1+\Sigma_{3/2}\Sigma_{3/2})\ ,
}
}
\eq{
\spl{
\int dz\Sigma{\cal O}=&\int dx^3\sqrt{g}(f^i\phi^iA^jA^j+\frac{1}{2}\psi^i\psi^iA^jA^j+B^iA^i\phi^j\phi^j
+\frac{1}{2}\lambda^i\lambda^i\phi^j\phi^j+2\psi^i\lambda^j\phi^iA^j)\\
=&\int dx^3\sqrt{g}({\cal O}_2\Sigma_1+\Sigma_2{\cal O}_1+2{\cal O}_{3/2}\Sigma_{3/2})\ ,
}
}
where we defined
\eq{
\spl{
&\Sigma_1=A^iA^i\ ,~~{\cal O}_1=\phi^i\phi^i ,~~\Sigma_{3/2}=A^i\lambda^i~,~~{\cal O}_{3/2}=\phi^i\psi^i\ ,
\\
&\Sigma_2=B^iA^i+\frac{1}{2}\lambda^i\lambda^i\ ,~~{\cal O}_2=f^i\phi^i+\frac{1}{2}\psi^i\psi^i\ ,
}
}
with the lower indices labeling the dimension of the single-trace operators.

With the above preparation the second factor of (\ref{deltaF}) at large $N$ is
\eq{
\spl{
&\int D\Sigma \exp\Big[\frac{1}{2\lambda}\int dz\Sigma^2+\frac{1}{2}\Big\langle\Big(\int dz\Sigma{\cal O}\Big)^2\Big\rangle_0\Big]
\\
=&\int D\Sigma \exp\Big[\frac{1}{\lambda}\int dx^3\sqrt{g}(\Sigma_2\Sigma_1+\Sigma_{3/2}\Sigma_{3/2})
\\
&+\frac{1}{2}\Big\langle\Big(\int dx^3\sqrt{g}({\cal O}_2\Sigma_1+\Sigma_2{\cal O}_1
+2{\cal O}_{3/2}\Sigma_{3/2})\Big)^2\Big\rangle_0\Big]
\\
=&\int D\Sigma \exp\Big[\frac{1}{\lambda}\int dV(\Sigma_2\Sigma_1+\Sigma_{3/2}\Sigma_{3/2})
\\
&+\frac{1}{2}\int\int dVdV'\Big(\Sigma_1(x)\Sigma_1(x')\Big\langle{\cal O}_2(x){\cal O}_2(x')\Big\rangle_0
+\Sigma_2(x)\Sigma_2(x')\Big\langle{\cal O}_1(x){\cal O}_1(x')\Big\rangle_0
\\
&+4\Sigma_{3/2}(x)\Sigma_{3/2}(x')\Big\langle{\cal O}_{3/2}(x){\cal O}_{3/2}(x')\Big\rangle_0\Big)\Big]\ ,
}
}
where $dV\equiv dx^3\sqrt{g}$, and we dropped vanishing terms in the two-point function to reach the last line.

The integral in \eqref{deltaF} then becomes gaussian, which integrates to give
\eq{
\frac{Z}{Z_0}=\frac{\det\Big(\mathds{1}+2\lambda\langle{\cal O}_{3/2}{\cal O}_{3/2}\rangle_0\Big)}
{\Big\{\det\Big(\frac{\lambda}{2}\langle{\cal O}_2{\cal O}_2\rangle_0\Big)\det\Big(\frac{\lambda}{2}
\langle{\cal O}_1{\cal O}_1\rangle_0\Big)\det\Big(\mathds{1}
-(\frac{\lambda}{4}\langle{\cal O}_2{\cal O}_2\rangle_0)^{-1}(\frac{\lambda}{4}
\langle{\cal O}_1{\cal O}_1\rangle_0)^{-1}\Big)\Big\}^{\frac{1}{2}}}\ .
}

At $\lambda\rightarrow\infty$, the change of the free energy compared to the free theory is
\eq{
\spl{
\Delta F=-\log\frac{Z}{Z_0}=&-{\rm tr}\log\Big(2\langle{\cal O}_{3/2}{\cal O}_{3/2}\rangle_0\Big)
+\frac{1}{2}{\rm tr}\log\Big(\frac{1}{2}\langle{\cal O}_2{\cal O}_2\rangle_0\Big)
\\
&+\frac{1}{2}{\rm tr}\log\Big(\frac{1}{2}\langle{\cal O}_1{\cal O}_1\rangle_0\Big)\ .
}\label{DF}
}
The two-point functions $\langle{\cal O}_1{\cal O}_1\rangle_0$ and $\langle{\cal O}_2{\cal O}_2\rangle_0$ 
can be expanded in terms of scalar harmonics on $S^3$ \cite{Gubser:2002vv}
\eq{
\langle{\cal O}_{\Delta}(x){\cal O}_{\Delta}(x')\rangle_0=\sum_{\ell m}g_{\ell}^{\Delta}Y^*_{\ell m}(x)Y_{\ell m}(x')\ ,
}
where $g_{\ell}^{\Delta}$ is given by
\eq{
g_{\ell}^{\Delta}=R^{3-2\Delta}\pi^{\ft32}2^{3-\Delta}\frac{\Gamma(\frac32-\Delta)}{\Gamma (\Delta)}\frac{\Gamma(\ell+\Delta)}{(3+\ell-\Delta)}\ .
}
Since the harmonics satisfy orthonormal relations, we have
\eq{
\int\sqrt{g}d^3y\langle{\cal O}_{2}(x){\cal O}_{2}(y)\rangle_0 \langle{\cal O}_{1}(y){\cal O}_{1}(x')\rangle_0
=\sum_{\ell m}g^{\Delta=2}_{\ell}g^{\Delta=1}_{\ell}Y^*_{\ell m}(x)Y_{\ell m}(x')\ .
}
It is straightforward to see that $g^{\Delta=2}_{\ell}g^{\Delta=1}$ is independent of $\ell$, and therefore according 
to \cite{Gubser:2002vv}, ${\rm tr}\log\langle{\cal O}_2{\cal O}_2\rangle_0+{\rm tr}\log\langle{\cal O}_1{\cal O}_1\rangle_0$ 
does not contribute to $\Delta F$.

Similarly, for fermionic two-point function, it is shown in \cite{Klebanov:2011gs} that 
${\rm tr}\log\langle{\cal O}_{3/2}{\cal O}_{3/2}\rangle_0$ is also zero. Therefore, in the IR there is no modification 
to the free energy given by the double-trace deformation.

When $\lambda$ is small, one can apply perturbation theory to compute $\Delta F$ induced by the deformation. As shown in \cite{Klebanov:2011gs} the change of free energy caused by the deformation is proportional to the beta function of the deformation coupling. The deformation appearing here is exactly marginal in the $N\rightarrow\infty$ limit, which implies that the beta function of the coupling constant 
is suppressed by $1/N$. Thus, at small coupling it can also be seen that the deformation does not affect the ${\cal O}(N^0)$ free energy. 
In summary, although we have not computed the free energy
of the deformed theory for arbitrary $\lambda$, the vanishing of $\Delta F$ at ${\cal O}(N^0)$ in both the strong and weak coupling limits provides strong evidence that $\Delta F$ does not receive ${\cal O}(N^0)$ contribution from the supersymmetric double-trace deformation, which is exactly marginal in the $N\rightarrow\infty$ limit.

\section{Conclusions}

We have carried out a one loop test of the conjectured dualities between Konstein-Vasiliev HS theories in $AdS_4$ with $S^3$ and $S^1_\beta\times S^2$ boundaries.  These theories are based on the HS algebras  $hu(m;n|4)$, $ho(m;n|4)$ and $husp(m;n|4)$ which contain $u(m) \oplus u(n)$, $o(m)\oplus o(n)$ and $usp(m)\oplus usp(n)$ as bosonic subalgebras. Generically these HS algebras can be interpreted as infinite dimensional supersymmetry algebras and they do not contain the extended $AdS_4$ superalgebra $OSp({\cal N}|4)$ as a subalgebra. They do so only in the special case of $m=n= 2^{\frac{\cal N}{2}-1}$ for even ${\cal N}$ or $2^{({\cal N}-1)/2}$ for odd ${\cal N}$. Our results for the free energies extend previous ones \cite{Giombi:2013fka,Giombi:2014iua,Giombi:2014yra} by inclusion of fermionic bulk degrees of freedom.  In computing the one loop free energies of bosonic and fermionic HS fields in $AdS_4$ with $S^3$ boundary, we have adopted the modified spectral zeta function method suggested by \cite{Bae:2016rgm}, thereby reproducing the one loop free energy for bosonic HS fields in a much simpler way without the ambiguities encountered in \cite{Giombi:2013fka,Giombi:2014iua}. We also find that the total one loop free energy of an infinite tower of bulk fermionic fields vanishes. 

Matching the bulk fields with boundary operators suggests that the possible CFT duals of Konstein-Vasiliev theories based on $hu(m;n|4)$, $ho(m;n|4)$ and $husp(m;n|4)$, and subject to HS symmetry preserving boundary conditions, are respectively the $U(N)$, $O(N)$ and $USp(N)$ singlet sectors of free scalars and free fermions vector representations of the bosonic subalgebras conformally coupled to $S^3$. We find that  the free energy of the HS theory may match with that of the free CFT only when the bulk theories are $hu(m;0|4)$, $ho(m;0|4)$, $husp(m;0|4)$ Konstein-Vasiliev theories, and with identifications $G^{-1}_N=\gamma(N+\Delta N)$ with suitable integers $\Delta N$. These are generalized type A theories with bosonic scalars on  the boundary and bosonic bulk HS fields containing even parity scalars.
Thus, in particular, the free energies for  generalized type B models with fermions on the $S^3$ boundary and bosonic HS fields including odd parity scalar fields do not match. The mismatch in the case of $m=0, n=1$ corresponding to the simplest type B model has already been noted in \cite{Giombi:2013fka} where 
the one loop free energy  $F^{(1)}=-\zeta(3)/(8\pi^2)$ obtained in the bulk does not agree with the free energy of Dirac fermions on $S^3$ boundary. 
We have also calculated the free energies of Konstein-Vasiliev theories in $AdS_4$ with $S^1_{\beta}\times S^2$ boundary. In this case, we find that the free energies of all three families of Konstein-Vasiliev theories match those of the conjectured dual free CFTs.

Turning to the problem of mismatch in free energies of type B model and its conjectured dual, 
one may have to take into account the issue of how to impose the $O(N)$ invariance condition
on the CFT side. A natural way of implementing it is to gauge the $O(N)$ symmetry by means 
of vector gauge field with 
level $k$ Chern-Simons kinetic term. This term breaks parity but the result 
for the free energy of the parity invariant model can be obtained in a limit 
in which the CS gauge field decouples. It has been suggested in \cite{Giombi:2013fka} that 
as the fermions coupled to CS on the boundary give rise to a shift in the level $k$, it may not be justified to obtain the result for parity-preserving case 
by a naive subtraction of CS contribution from the free energy on the CFT side.
However, one expects that this effect becomes irrelevant in the decoupling limit 
in which $k\to \infty$. In fact, we have examined the procedure of decoupling CS in the large $k$ limit by evaluating the $S^3$ free energies for ABJ model based on $U(N)_k\times U(1)_{-k}$ \cite{Kapustin:2009kz,Awata:2012jb} and a few ${\cal N}=3$ CS matter theories in which the matter sector consists of fundamental hypermultiplets \cite{Marino:2011nm,Gulotta:2012yd,Mezei:2013gqa}. After subtracting the contribution from pure CS term, we indeed obtain the free energies of free vector models. Therefore, the puzzle of free energy mismatch in type B 
remains unresolved and its solution requires deeper understanding of HS/vector model holography. In this context, it has been suggested by \cite{Chang:2012kt} and explored further in \cite{Hirano:2015yha} that the vector-like limit of  ABJ model based on $U(N)_k\times U(M)_{-k}$ is given by
\be
N,\,k\rightarrow\infty \quad {\rm with }\quad \lambda\equiv 
\frac{N}k\quad {\rm and} \quad M\quad  {\rm finite}\ . 
\label{vector-limit}
\ee
 In this limit, the ABJ theory effectively behaves like a ${\cal N}=6$ CS gauged vector model with $U(M)$ flavor symmetry \cite{Chang:2012kt}. Its bulk dual is conjectured to be the parity violating ${\cal N}=6$ $U(M)$ gauged Vasiliev theory \cite{Chang:2012kt}. The parity violating angle $\theta_0$ is conjectured to be related to the CFT t'Hooft coupling by $\theta_0=\pi\lambda/2$ \cite{Chang:2012kt}
 \footnote{ Besides the Newton constant which is small in the limit described above, there is also a bulk t'Hooft coupling $g^2_{\rm bulk} M\sim M/N\ll1$. String theory emerges when $M/N\sim 1$. Due to strong interactions, the HS particles form $U(M)$ singlet states which are described by the color neutral string states. Since the M theory circle $R_{11}\sim (M/k^5)^{1/6}$ shrinks and $\sqrt{\alpha'}/R_{\rm AdS}\sim (k/M)^{1/4}\rightarrow \infty$, this is type IIA string in the high energy limit. The ${\cal N}=6$ parity violating $U(M)$ gauged Vasiliev theory can be perceived as a deconfinement phase of type IIA string when $M/N\ll 1$, in which the string states fragment into HS particles colored under $U(M)$ \cite{Chang:2012kt}.  }.

Turning to the question of free energy in the parity invariant HS theory, we may first keep $\lambda$ finite, and consider the limit $\lambda \to 0$ that is required for the parity invariant limit at the end
\footnote{There are subtleties regarding the $\lambda\rightarrow 0$ limit having to do with the subtraction of the free energy coming from the CS term, which may correspond to subtraction of an open string sector in the bulk \cite{Chang:2012kt}.} 
%
%
Different from the  parity preserving HS theories, in the ${\cal N}=6$ parity violating HS theory a mixed boundary condition needs to be imposed on the bulk spin-1 gauged field in order to preserve the ${\cal N}=6$ supersymmetry \cite{Chang:2012kt}. 
%
%
The effect of the mixed spin-1 boundary condition was mimicked by introducing an $N$-dependent anomalous dimension for the bulk spin-1 gauge field, which is responsible for the $\log N$ term in the one loop free energy of the bulk theory. The bulk one loop free energy is then compared with the free energy of ABJ theory in the vector-like limit \eqref{vector-limit}, and with the free energy of pure $U(M)$ CS subtracted. Matching of the $\log N$ terms present in the free energies on both sides leads to the identification \cite{Hirano:2015yha}
\be
G_N=\frac{\gamma}{N}\frac{\pi\lambda}{\sin(\pi \lambda)}\ .
\ee
On the other hand, an exact expression for $G_N$ has been obtained from correlation function for two stress tensors on the CFT side in \cite{Honda:2015sxa}. Comparing the relevant terms in these expressions for $G_N$ one deduces that $\gamma=2/\pi$. 
Assuming the stated value of $\gamma$, in the limit $\lambda\rightarrow 0$, required for obtaining the parity invariant HS theory, one finds the relation $G_N=2/(N\pi)$ which differs from the one that appears in the HS/free vector model holography by a factor of $\pi$.
This is due to the fact that, while we assume that $F_{\rm bulk}^{(0)}=F_{\rm CFT}^{(0)}$ in the HS/free vector model holography, the example of HS/ABJ holography seems to suggest that $F^{(0)}_{\rm bulk}=F_{\rm CFT}^{(0)}/\gamma$. 
The above approach may seem to resolve the free energy problem in type B model, however 
what is missing in this picture is that a bulk computation of the AdS energy for the vector field to justify this value of $\gamma$. Furthermore, the beyond $\log N$ dependence, the terms of higher order in $1/N$ have not been compared in the matching of the free energies. These issues clearly deserves further study.

Another interesting future direction is to consider HS/free matrix model holography. 
In this case, the corresponding bulk HS theory contains infinitely many massive HS fields in addition to the usual massless ones. Recently, a preliminary one loop test of HS/free matrix model holography was carried out in \cite{Bae:2016rgm}. A dual pair considered in \cite{Bae:2016rgm} consists of a free scalar field, namely the bosonic singleton, namely  Rac, in the adjoint representation of $SU(N)$ and a HS theory in $AdS_4$ whose spectrum can be constructed from the two, three and four-fold tensor products of  the Rac. The bulk fields are dual to the single-trace of product of multiple Rac's. The one loop free energies of the bulk fields belonging to the first few Regge trajectories were computed in \cite{Bae:2016rgm}. The one loop free energy of the first trajectory comprised of massless HS fields is equal to that of a real conformally coupled scalar, however, such feature ceases to exist for higher trajectories. It is possible that after summing over all trajectories the total bulk free energy may possess a nice property. But such a difficult task has not been completed. It is also possible that supersymmetry may provide simplifications, as we recall that in $AdS_5$, the long multiplet of $SU(2,2|4)$ gives rise to vanishing one loop free energy \cite{Beccaria:2014xda}. It should be noted that the matrix phase of ABJ model based on $U(N)_k\times U(M)_{-k}$ with $M\sim N$ has conserved HS currents emerging in the limit $\lambda\rightarrow 0$, which implies the presence of massless HS particles in the spectrum of type IIA string. Thus, in the regime 
\be
M\sim N\ ,\qquad  \lambda = N/k\rightarrow 0\ , 
\label{R2}
\ee
the duality between IIA string on $AdS_4\times\mathbb{CP}^3$ and ABJ theory may provide an example of HS/free matrix model duality \cite{Chang:2012kt} if the contribution from CS term in the CFT can be simply subtracted. For the string theory interpretation of this limit, we refer the reader to \cite{Aharony:2008ug}. The point we wish to stress here is that there are two regimes of type IIA string theory on $AdS_4 \times \mathbb{CP}^3$ which remarkably give two different supersymmetric HS theories one of which is expected to be dual to a vector model, and the other to a matrix model on the boundary of $AdS_4$, and that the puzzle we have encountered in the one loop test of holography by computing the free energies in the case of vector model remains to be investigated thoroughly in the case of  matrix model.

A complete matching of the free energies on both sides requires the knowledge of $F^{(0)}_{\rm bulk}$ which can only be computed from the full action for HS theory. There exists an action that takes the form of a Chern-Simons action in a generalized spacetime of  form ${\cal M}_9= {\cal X}_5\times {\cal Z}_4$ where 
${\cal Z}_4$ is a twistor space with no boundary, and the spacetime ${\cal M}_4$ resides on an open region of 
the boundary of ${\cal X}_5$ \cite{Boulanger:2015kfa}.  The action contains Lagrange multiplier master fields but they do not 
propagate to produce unwanted degrees of freedom. What remains to be done is to add suitable HS 
invariant deformations  that reside on the boundary of ${\cal M}_9$, which are highly restricted 
and for which candidates have been proposed \cite{Boulanger:2015kfa}, and to construct a boundary action that 
resides on the boundary of asymptotically $AdS_4$ spacetime ${\cal M}_4$ which has not been constructed so far.  
These are needed for obtaining the field equations through an appropriate variational principle, and once they are constructed, the full action can be quantized in a path integral approach and the Feynman rules can be derived, even though the action does not have  
the traditional form consisting an infinite sum of Einstein-Hilbert term and powers of curvature tensors and their derivatives. It remains to be seen whether the result for the one loop free energy computed in this fashion agrees with that obtained under the assumption that the quadratic action for the HS fluctuations around $AdS_4$ has the standard Fronsdal form with two derivatives. In particular, it would be interesting to determine if the mismatch in the free energies encountered in the type B and ordinary supersymmetric HS theories and their conjectured duals may find a resolution in a computation based on the action discussed above. 


\subsection*{Acknowledgements}

We thank  S. Giombi, I. Klebanov, E. Skvortsov, P. Sundell, A. Tseytlin, M. Vasiliev and X. Yin 
for discussions. Y.P. would like to thank Nordita Institute, Beijing Normal University and Sun Yet-sen University for hospitality during various stages of this work. Y.P and E.S. would also like to thank Munich Institute for Astro-and Particle Physics (MIAPP) for providing a wonderful work environment. The work of E.S. is supported in part by NSF grant PHY-1214344. Y.P. is  supported by Alexander von Humboldt fellowship.


\appendix



\section{Comparison of
$\zeta_{(\Delta, s)}(z)$ with $\widetilde{\zeta}_{(\Delta, s)}(z)$}

In this section, we will show that the alternate spectral zeta function is physically equivalent to
the original spectral zeta in computing the one loop free energy of HS fields.  

\subsection{Bosonic case}

For bosonic HS fields, the physical equivalence of alternate spectral zeta function and the original spectral
zeta function has been studied in \cite{Bae:2016rgm} in the case of summing over all integer spins. The crucial point is that for a given HS field labeled by $(\Delta,\,s)$, the difference
between the alternate spectral zeta function and the original zeta function can be expressed as a contour integral encircling $\beta=0$ \cite{Bae:2016rgm}
\bea
&&\widetilde{\zeta}^{B}_{(\Delta, s)}(z)-\zeta^B_{(\Delta, s)}(z)=\frac{1}{3}\left(s+\frac{1}{2}\right)\nu^2\left[\frac{1}{6}\nu^2-\left(s+\frac{1}{2}\right)^2\right]\,
\\
&=&\frac{z}{2\pi i}\oint d\beta\frac{2\sinh^3\frac{\beta}{2}}{\beta^3}
\left(\frac{8}{3\beta^2}+\frac{2}{\sinh^2\frac{\beta}{2}}-\frac{1}{3}
+4\partial^2_\alpha\right)\chi_{\Delta,s}(\beta,\alpha)\Big|_{\alpha=0}+{\cal O}(z^2)\ .
\nn
\eea
It has been shown in \cite{Bae:2016rgm} that upon summing over all integer spins, the contour integral vanishes. We have also checked that this is also true for summing over all even spins or odd spins separately.

\subsection{Fermionic case}

For fermionic HS fields, we will elaborate on the physical equivalence of alternate spectral zeta function and the original spectral
zeta function which has not been studied elsewhere.
For a fermionic HS field labeled by $(\Delta,\,s)$, the alternate spectral zeta function can be computed exactly
\eq{
\spl{
\widetilde{\zeta}^F_{(\Delta, s)}(z)=&(2s+1)\left(\frac{1}{32}-\frac{s(s+1)}{24}\right)\frac{1}{\Gamma(2z)}
\int^\infty_0d\beta\beta^{2z-1}e^{-\nu\beta}\frac{1}{\sinh^2\frac{\beta}{2}}
\\
&+\frac{2s+1}{16}\frac{1}{\Gamma(2z)}\int^\infty_0d\beta\beta^{2z-1}
e^{-\nu\beta}\frac{1}{\sinh^4\frac{\beta}{2}}
\\
=& \frac{2s+1}{24}\big[\nu\left( (2s+1)^2-4\nu^2\right)\zeta(2z,\nu)+4\zeta(2z-3,\nu)-12\nu\zeta(2z-2,\nu)
\\
&+\left(12\nu^2-4s(s+1)-1\right)\zeta(2z-1,\nu)\big]\ ,
}
}
from which we see that $\widetilde{\zeta}^F_{(\Delta, s)}(0)$ matches $\zeta^F_{(\Delta, s)}(0)$. The latter takes the form
\be
\zeta^B_{(\Delta, s)}(0)&=&\frac{s+\frac{1}{2}}{6} \left[\frac{\nu^4}{2} - (s+\frac{1}{2})^2 \nu^2\right]-\frac{1}{3} (2s+1) \left[\frac{7}{1920} + \frac{(s + \frac{1}{2})^2}{48}\right]\ .
\label{zetab0}
\ee
Next, we compute the first derivative of $\widetilde{\zeta}^F_{(\Delta, s)}(z)$ at $z=0$, which is given by
\be
\widetilde{\zeta}^{F\prime}_{(\Delta, s)}(0) &=&\frac{2s+1}{12}\big[\nu\left((2s+1)^2-4\nu^2\right)\zeta'(0,\nu)+4\zeta'(-3,\nu)-12\nu\zeta'(-2,\nu)
\nn\\
&&+(12\nu^2-4s(s+1)-1)\zeta'(-1,\nu)\big]\ .
\ee
After some algebra, we obtain the difference between $\widetilde{\zeta}^{F\prime}_{(\Delta, s)}(0)$ and $\zeta^{F\prime}_{(\Delta, s)}(0)$
\be
\widetilde{\zeta}^{F\prime}_{(\Delta, s)}(0)-\zeta^{F\prime}_{(\Delta,s)}(0)=-\frac{1}{24}(2s+1)^3\nu^2+\frac{2s+1}{9}\nu^4\ .
\ee
This can again be converted to a contour integral of $\beta$ circling $\beta=0$
\be
\widetilde{\zeta}^{F\prime}_{(\Delta, s)}(0)-\zeta^{F\prime}_{(\Delta,s)}(0)={2\pi i}\oint d\beta\frac{2\sinh^3\frac{\beta}{2}}{\beta^3}\Big(\frac{32}{3\beta^2}+\frac{2}{\sinh^2\frac{\beta}{2}}-\frac{1}{3}+4\partial^2_\alpha\Big)\chi_{\Delta,s}(\beta,\alpha)\ .
\ee
From \eqref{c9}, one can see that the total character of fermionic sector including the contributions of all physical fermionic higher fields and their ghosts gives rise to an even function of $\beta$ which has vanishing contour integral. Therefore, we have shown that in computing the one loop free energy of the whole fermionic sector, the alternate spectral zeta function is physically equivalent to the original one.

\newpage


{}

\end{document}